\documentclass[12pt]{article}

\usepackage{latexsym}
\usepackage{amssymb,amsfonts,amsmath}
\usepackage{graphicx} 
\usepackage{indentfirst}
\usepackage{bbm}
\usepackage{amssymb}
\usepackage{verbatim}
\usepackage{amsmath, amsthm,amssymb}
\usepackage{mathrsfs}
\usepackage{hyperref}
\usepackage{amsfonts}
\usepackage{dsfont}

\parskip=\medskipamount

\newcommand {\cA}{{\cal A}}

\newcommand {\cC}{{\cal C}}
\newcommand {\cD}{{\cal D}}
\newcommand {\cE}{{\cal E}}

\newcommand {\cJ}{{\cal J}}

\newcommand {\cL}{{\cal L}}
\newcommand {\cM}{{\cal M}}
\newcommand {\cN}{{\cal N}}

\newcommand {\cR}{{\cal R}}
\newcommand {\cS}{{\cal S}}

\newcommand {\cV}{{\cal V}}
\newcommand {\cW}{{\cal W}}

\def\a{\alpha}

\def\b{\beta}

\def\d{\delta}

\def\g{\gamma}
\def\G{\Gamma}

\def\k{\kappa}
\def\l{\lambda}
\def\m{\mu}

\def\o{\omega}

\def\q{\theta}
\def\r{\rho}
\def\s{\sigma}

\def\x{\xi}
\def\z{\zeta}
\def\D{\Delta}
\def\F{\Phi}
\def\J{\Psi}
\def\L{\Lambda}
\def\O{\Omega}

\def\S{\Sigma}

\def\tr{{\rm tr}}
\def\rd{{\rm d}}
\def\ri{{\rm i}}
\def\re{{\rm e}}

\newcommand{\ve}{\varepsilon}                            
\newcommand{\cDB}{{\bar\cD}}                            
\newcommand{\DB}{\bar{D}}

\newcommand{\ab}{{\a\b}}

\newcommand{\pa}{\partial}                           
\newcommand{\hf}{\frac12}

\newcommand{\vf}{\varphi}

\newcommand{\be}{\begin{equation}}
\newcommand{\ee}{\end{equation}}
\newcommand{\bea}{\begin{eqnarray}}
\newcommand{\eea}{\end{eqnarray}}
\newcommand{\non}{\nonumber}
\newcommand{\1}{{\underline{1}}}
\newcommand{\2}{{\underline{2}}}

\newcommand{\bm}[1]{\mbox{\boldmath$#1$}}

\def\double #1{#1{\hbox{\kern-2pt $#1$}}}

\newcommand{\qb}{{\bar{\theta}}}

\newif\ifdtup

\def\de{{\nabla}}                                         

\newcommand{\bsubeq}{\begin{subequations}}
\newcommand{\esubeq}{\end{subequations}}

\newcommand{\mub}{{{\bar{\mu}}}}

\numberwithin{equation}{section}

\begin{document}

\begin{titlepage}
\begin{flushright}
July, 2018 \\
\end{flushright}
\vspace{5mm}

\begin{center}
{\Large \bf 
  \mbox{$\bm{\cN=2}$} 
supersymmetric higher spin gauge theories
and current multiplets
in three dimensions}
\\ 
\end{center}

\begin{center}

{\bf
Jessica Hutomo, Sergei M. Kuzenko and Daniel Ogburn } \\
\vspace{5mm}

\footnotesize{
{\it Department of Physics M013, The University of Western Australia\\
35 Stirling Highway, Crawley W.A. 6009, Australia}}  
~\\

\vspace{2mm}
~\\
\texttt{jessica.hutomo@research.uwa.edu.au, sergei.kuzenko@uwa.edu.au,
daniel.x.ogburn@gmail.com}\\
\vspace{2mm}

\end{center}

\begin{abstract}
\baselineskip=14pt
We describe several families of primary linear supermultiplets coupled 
to three-dimensional ${\cal N}=2$ conformal supergravity 
and use them to construct topological $BF$-type terms.
We introduce conformal higher-spin gauge superfields 
and associate with them Chern-Simons-type actions that are constructed 
as an extension of the linearised action for ${\cal N}=2$ conformal supergravity. 
These actions possess gauge 
and super-Weyl invariance in any conformally flat superspace
and involve a higher-spin generalisation of the linearised 
${\cal N}=2$ super-Cotton tensor.
For massless higher-spin supermultiplets in (1,1) anti-de Sitter (AdS) superspace,
we propose two  off-shell Lagrangian gauge formulations, which are related 
to each other by a dually transformation. Making use of these massless 
theories allows us to formulate consistent higher-spin supercurrent 
multiplets in (1,1) AdS superspace. Explicit examples of such supercurrent
multiplets are provided for models of massive chiral supermultiplets. 
Off-shell formulations for massive higher-spin supermultiplets
in (1,1) AdS superspace are proposed.
\end{abstract}
\vspace{5mm}

\begin{flushright}
{\it Dedicated to Professor Ioseph L. Buchbinder}\\
{\it on the occasion of his 70th birthday}
\end{flushright}

\vfill

\vfill
\end{titlepage}

\newpage
\renewcommand{\thefootnote}{\arabic{footnote}}
\setcounter{footnote}{0}

\tableofcontents{}
\vspace{1cm}
\bigskip\hrule

\allowdisplaybreaks

\section{Introduction}

In four spacetime dimensions (4D), there exist two off-shell formulations 
for pure $\cN=1$ anti-de Sitter (AdS) supergravity: minimal (see, e.g., \cite{GGRS,BK} 
for pedagogical reviews) and non-minimal \cite{BK12}.\footnote{It was believed for almost thirty 
years that there is no non-minimal formulation for $\cN=1$ AdS supergravity
\cite{GGRS}. However, such a formulation was constructed in \cite{BK12}.}
These supergravity theories are related to each other by 
a superfield duality transformation \cite{BK12} and 
possess a single maximally supersymmetric solution, 
the famous  $\cN=1$ AdS superspace  \cite{Keck,Zumino77,IS}, 
which is the simplest member of the family of $\cN$-extended AdS superspaces
\bea
{\rm AdS}^{4|4\cN} = \frac{{\rm OSp}(2|4)}{{\rm SO}(3,1) \times {\rm SO} (\cN)}~.
\eea
These supergravity theories are also intimately related to the two
dually equivalent series of massless gauge supermultiplets 
of half-integer superspin $s+\hf \geq \frac 32$
(describing two ordinary massless spin-$(s+\hf)$ and spin-$(s+1)$ fields on the mass shell) 
in AdS${}_4$, which were   proposed in \cite{KS94} 
as a natural extension of the formulations 
 in Minkowski space constructed earlier in \cite{KSP,KS}. 
 Specifically, for the lowest superspin value corresponding to $s=1$, one series yields the linearised action for
minimal  AdS supergravity, while the other leads to   
 the linearised non-minimal  AdS supergravity.

In the 3D case, the AdS  group is reducible, 
$$\rm SO(2,2) \cong \Big( SL(2, {\mathbb R}) \times SL( 2, {\mathbb R}) \Big)/{\mathbb Z}_2~,$$ 
and so are its simplest supersymmetric extensions,  
${\rm OSp} (p|2; {\mathbb R} ) \times  {\rm OSp} (q|2; {\mathbb R} )$.
This implies that  $\cN$-extended AdS supergravity exists in several incarnations  \cite{AT}. 
These are known as the  $(p,q)$ AdS supergravity theories,
where the  non-negative integers $p \geq q$ are such that $\cN=p+q$.   
For any allowed values of $p$ and $q$, 
the pure  $(p,q)$ AdS supergravity was constructed in \cite {AT}
as a Chern-Simons theory with the gauge group
 ${\rm OSp} (p|2; {\mathbb R} ) \times  {\rm OSp} (q|2; {\mathbb R} )$.
The Chern-Simons construction is not particularly useful when one is interested 
in coupling AdS supergravity to supersymmetric matter. 
This is one of the reasons why off-shell formulations for 
3D $\cN$-extended conformal supergravity 
have been developed \cite{HIPT,KLT-M11,BKNT-M1}.

Within the off-shell supergravity framework of  \cite{KLT-M11}, $(p,q)$ AdS superspace 
\bea
{\rm AdS}^{(3|p,q)} = \frac{ {\rm OSp} (p|2; {\mathbb R} ) \times  {\rm OSp} (q|2; {\mathbb R} ) } 
{ {\rm SL}( 2, {\mathbb R}) \times {\rm SO}(p) \times {\rm SO}(q)}
\eea
originates as  a maximally symmetric conformally flat supergeometry 
with covariantly constant torsion and curvature 
generated by a  tensor $S^{IJ}= S^{JI}$  \cite{KLT-M12}, 
with the SO$(\cN)$ indices $I, J$ 
taking values from 1 to $\cN$.  
It turns out that the symmetric matrix $S=(S^{IJ})$ 
is nonsingular, and the parameters $p$ and $q= \cN-p$ 
determine its  signature. The ordinary AdS space 
\bea
{\rm AdS}_{3} = \frac{ \rm SL( 2, {\mathbb R})  \times  SL( 2, {\mathbb R})  } 
{\rm SL( 2, {\mathbb R}) }
\eea
is the bosonic body of ${\rm AdS}^{(3|p,q)} $. 
The curvature of ${\rm AdS}_{3} $ is proportional to 
$\tr (S^2)$. The Killing vector fields of ${\rm AdS}^{(3|p,q)}$  
can be shown to generate the isometry group
${\rm OSp} (p|2; {\mathbb R} ) \times  {\rm OSp} (q|2; {\mathbb R} )$, 
see \cite{KLT-M12} for the technical details.

The 3D $\cN=2$ supersymmetry is a natural cousin of the 4D $\cN=1$ one.
This is the lowest value of $\cN$ for which there are  
at least two inequivalent AdS superspaces, 
 ${\rm AdS}^{(3|1,1)}$ and ${\rm AdS}^{(3|2,0)}$, which were 
thoroughly studied in \cite{KT-M11}.
The former is the 3D counterpart of the 4D $\cN=1$ AdS superspace, while 
the latter has no 4D analogue.  
The superspaces  ${\rm AdS}^{(3|1,1)}$ and ${\rm AdS}^{(3|2,0)}$ 
are maximally symmetric solutions of the known  off-shell $\cN=2$ 
AdS supergravity theories  presented in \cite{KT-M11}. 
${\rm AdS}^{(3|1,1)}$ is the unique maximally symmetric solution 
of the two dually equivalent  (1,1) AdS supergravity theories,
minimal and non-minimal ones.
${\rm AdS}^{(3|2,0)}$ is the unique maximally symmetric solution 
of the (2,0) AdS supergravity,  which was originally formulated in \cite{HIPT} 
in the component setting. The early superspace descriptions of
 the minimal (1,1) supergravity were given in \cite{ZP,NG}.

Since there are three off-shell $\cN=2$ AdS supergravity theories \cite{KT-M11}, 
one might expect existence of three  series of massless 
higher-spin gauge supermultiplets.  
In this paper we present two series of massless higher-spin actions,
 which are associated with the minimal and the non-minimal (1,1) AdS 
supergravity theories, respectively, and 
which generalise similar constructions in the super-Poincar\'e case \cite{KO}.
Off-shell higher-spin actions with (2,0) AdS supersymmetry 
will be described in a separate work \cite{HK2018}.

Similar to the pure gravity and simple supergravity theories in three dimensions, 
pure $\cN=2$ supergravity 
(massless superspin-3/2 multiplet) 
and its higher-spin extensions have no propagating degrees of freedom.
Nevertheless, there are at least two nontrivial applications of 
the massless higher-spin gauge supermultiplets.
Firstly, one can follow the pattern of topologically massive (super)gravity
\cite{DJT1,DJT2,DK,Deser} and construct massive higher-spin supermultiplets
by combining a massless 
action with a higher-spin 
extension of the action for linearised conformal supergravity. 
This has been achieved in \cite{KO} in the $\cN=2$ super-Poincar\'e case, 
and similar ideas have been implemented  in the frameworks 
of $\cN=1$ Poincar\'e and AdS supersymmetry  \cite{KT,KP}.
Secondly, making use of the   off-shell formulations for massless higher-spin 
supermultiplets in ${\rm AdS}_{3} $, one can define consistent higher-spin supercurrent 
multiplets in AdS superspace
(i.e. higher-spin extensions  of the supercurrent)
 that contain ordinary bosonic and fermionic conserved currents in ${\rm AdS}_{3} $. 
One can then look for explicit realisations of such 
higher-spin supercurrents in concrete supersymmetric theories in ${\rm AdS}_{3} $. 
Such a program in the 4D $\cN=1$ Poincar\'e and AdS supersymmetric cases 
has been described in a series of papers \cite{KMT,HK1,HK2,BHK}. 
Alternatively, one can develop a 3D extension of the approach 
advocated in \cite{BGK1,KKvU,BGK3} and based on the use of superfield 
Noether procedures \cite{Osborn,MSW}.

Before we turn to the main body of this work, a few comments are in order 
about maximally supersymmetric backgrounds 
in the off-shell $\cN=2$ supergravity theories, since the superspaces 
 ${\rm AdS}^{(3|1,1)}$ and ${\rm AdS}^{(3|2,0)}$ are special examples 
 of such supermanifolds.
The most general maximally supersymmetric backgrounds 
are characterised by several conditions \cite{KLRST-M,K15Corfu}
on the  torsion superfields $\cR$, $\cS$ and $\cC_a$, 
which determine the superspace geometry of $\cN=2$ conformal supergravity
(see section 2 for the technical details).
These requirements are as follows:
\begin{subequations}
\bea
\cR \cS&=&0~, \qquad \cR \,{\cC}_a = 0 ~, \\
{\cD}_A  {\cR}&=& 0~, \qquad
{\cD}_A  {\cS}= 0~,
\qquad \cD_\a \cC_b =0 \quad \Longrightarrow \quad
{\cD}_{a} {\cC}_b=
2\ve_{abc}{\cC}^c {\cS} ~.~~~
\eea
\end{subequations}
The (1,1) AdS superspace is singled out by 
the conditions $\cS=0$ and ${\cC}_a = 0$,
 with $\cR$ and its conjugate $\bar \cR$
having non-zero constant values.
The  (1,1) AdS superspace is characterised by the 
following algebra of covariant derivatives \cite{KT-M11}:
\begin{subequations} \label{(1,1)AdS}
\bea
\{\cD_\a,\cD_\b\}
&=&
-2\bar{\cR}
 (\g_a)_{\a\b} \ve^{abc}M_{bc} 
~,\qquad
\{\cDB_\a,\cDB_\b\}
=2{\cR}(\g_a)_{\a\b} \ve^{abc}M_{bc} ~,
\\
\{\cD_\a,\cDB_\b\}
&=&
-2 \ri (\g^c)_{\a\b}   \cD_c~, \\
{[}\cD_{a},\cD_\b{]}
&=&
\ri(\g_a)_{\b}{}^{\g}\bar{\cR}\cDB_{\g}
~,\\
{[}\cD_{a},\cDB_\b{]}
&=&
-\ri(\g_a)_\b{}^{\g}{\cR}\cD_{\g}~, \\
{[}\cD_a,\cD_b]{}
&=&-4  
\bar{\cR}\cR 
M_{ab}~,
\eea
\end{subequations}
where $M_{ab}$ denotes the Lorentz generator.
The (2,0) AdS superspace belongs to the family of  
all maximally supersymmetric backgrounds with $\cR=0$.
These backgrounds are characterised by the following  algebra of covariant derivatives \cite{KLRST-M,K15Corfu}:
\begin{subequations}
\bea
\{\cD_\a,\cD_\b\}
&=&
0
~,\qquad
\{\cDB_\a,\cDB_\b\}
=
0~,
\\
\{\cD_\a,\cDB_\b\}
&=&
-2 \ri (\g^c)_{\a\b} \Big(  \cD_c - 2\cS M_c
-\ri  \cC_{c} J \Big)
+4\ve_{\a\b}\Big( \cC^{c}M_{c}- \ri \cS J\Big)
~, \\
{[}\cD_{a},\cD_\b{]}
&=&
\ri\ve_{abc}(\g^b)_\b{}^{\g}\cC^c\cD_{\g}
+ (\g_a)_\b{}^\g \cS \cD_{\g}~, \\
{[}\cD_{a},\cDB_\b{]}
&=&
-\ri\ve_{abc}(\g^b)_\b{}^{\g}\cC^c\cDB_{\g}
+(\g_a)_\b{}^{\g}\cS \bar \cD_{\g}~, \\
{[}\cD_a,\cD_b]{}
&=&4  \ve_{abc}\Big( \cC^c \cC_d
+\d^c{}_d \cS^2
\Big)M^d ~.
\eea
\end{subequations}
Here $J$ is the generator of the $\cN=2$ $R$-symmetry group, ${\rm U(1)}_R$, 
and  $M^a:=\hf\ve^{abc}M_{bc}$.
The solution with ${\cC}_a =0$ and $\cS\neq 0$ corresponds to  (2,0) AdS superspace
\cite{KT-M11}. It may be shown that the ${\rm U(1)}_R$ connection is flat if and only if 
$\cS =0$ \cite{KLT-M11}. 
The non-vanishing ${\rm U(1)}_R$ curvature 
is the main reason why 
the structure of massless higher-spin gauge supermultiplets in (2,0) AdS superspace
\cite{HK2018}
considerably differs from their counterparts with (1,1) AdS supersymmetry.

This paper is organised as follows. 
In section 2, primary linear supermultiplets coupled to $\cN=2$
conformal supergravity are described  and then used to construct 
topological $BF$-type terms. Given a positive integer $n>0$,
we introduce a conformal gauge superfield ${\mathfrak H}_{\a(n)}$ and show that, 
for every conformally flat superspace, there exists 
a unique primary  gauge-invariant descendant 
${\mathfrak W}_{\a(n)} ({\mathfrak H})$ 
of ${\mathfrak H}_{\a(n)}$ with the properties \eqref{2.25}.
In terms of ${\mathfrak H}_{\a(n)}$ and ${\mathfrak W}_{\a(n)}({\mathfrak H})$
we construct  a higher-spin extension 
of the linearised action for $\cN=2$ conformal supergravity.
Section 3 provides a brief summary of the key results
concerning the (1,1) AdS superspace and superfield representations
of the corresponding isometry group. In sections 4 and 5,
we present two dually equivalent  off-shell Lagrangian  formulations
for every massless higher-spin supermultiplet in (1,1) AdS superspace.
 Making use of these massless 
theories allows us to formulate, in section 6, 
consistent higher-spin supercurrent multiplets. 
Explicit examples of such supercurrents are provided in sections 7 and 8
for models described by chiral supermultiplets.
In section 9 we discuss several extensions of the constructions obtained.
The paper is concluded with two appendices.
Appendix A describes our notation, conventions and 
several important identities involving the spinor covariant derivatives 
of (1,1) AdS superspace. 
Appendix B describes the ${\cN=2} \rightarrow {\cN=1}$ superspace 
reduction  of the massless integer superspin 
model \eqref{long-action-int} in Minkowski superspace.


\section{Superconformal higher-spin multiplets}

Before presenting superconformal higher-spin multiplets, 
we  give a succinct review of 3D $\cN=2$ conformal supergravity 
following \cite{HIPT,KLT-M11}. There exists more general formulation 
for conformal supergravity  \cite{BKNT-M1}
known as the $\cN=2$ conformal superspace. For our purposes it suffices to 
use the formulation of \cite{KLT-M11}, which is obtained from the $\cN=2$ conformal 
superspace by partially fixing the gauge freedom.

\subsection{Conformal supergravity}

All known off-shell formulations for 3D $\cN=2$ supergravity \cite{KLT-M11,KT-M11}
can be realised in curved superspace $\cM^{3|4}$
 with  structure group  ${\rm SL}(2,{\mathbb{R}})\times {\rm U(1)}_R$, 
 where ${\rm SL}(2,{\mathbb{R}})$ and $ {\rm U(1)}_R$
 stand for the spin group and the $R$-symmetry group, respectively.
The superspace  is  parametrised by
local bosonic ($x^m$) and fermionic ($\q^\m, \bar \q_\m$)
coordinates  $z^{{M}}=(x^{m},\q^{\mu},{\bar \q}_{{\mu}})$,
where the  variables $\q^{\mu} $ and $\bar \q_{{\mu}}$
are related to each other by complex conjugation:
$\overline{\q^{\mu}}=\bar \q^{{\mu}}$.

The superspace covariant derivatives have the form
\bea
\cD_{{A}}=(\cD_{{a}}, \cD_{{\a}},\bar \cD^\a)
=E_{{A}}+\O_{{A}}+\ri \F_{{A}} J~.
\label{CovDev}
\eea
Here $E_A$ and $\O_A$ denote
 the inverse supervielbein and 
 the Lorentz connection, respectively,
\bea
E_A=E_A{}^M \frac{\pa}{\pa z^M}~,
\qquad \O_A=\hf\O_{A}{}^{bc} M_{bc}= -\O_{A}{}^b M_b
=\hf\O_{A}{}^{\b\g}M_{\b\g}~.
\eea
The Lorentz generators with two vector indices 
($M_{ab}= -M_{ba}$), with one vector index ($M_a$)
and with two spinor indices ($M_{\a\b} = M_{\b\a} $) 
are defined in Appendix \ref{AppendixA}.
The ${\rm U(1)}_R$ generator $J$  in \eqref{CovDev}
is defined to act on the covariant derivatives
as follows:
\bea
{[} J,\cD_{\a}{]}
=\cD_{\a}~,
\qquad
{[} J,\cDB^{\a}{]}
=-\cDB^\a~,
\qquad 
{[}J,\cD_a{]}=0~.
\eea

In order to describe $\cN=2$ conformal supergravity, the torsion 
has to obey the covariant constraints proposed in \cite{HIPT}.
Solving the constraints leads to the following
algebra of covariant derivatives  \cite{KLT-M11,KT-M11}
\bsubeq \label{algebra-final}
\bea
\{ \cD_\a , \cD_\b \} &=& - 4 \bar \cR M_{\a\b} \ , \\
\{ \cD_\a , {\bar \cD}_\b \} &=&
- 2 \ri (\g^c)_{\a\b} \cD_c 
- 2 \cC_{\a\b} J
- 4 \ri \ve_{\a\b} \cS J 
+ 4 \ri \cS M_{\a\b}
- 2 \ve_{\a\b} \cC^{\g\d} M_{\g\d} \ , \label{algebra-final-b} \\
{[}\cD_a , \cD_\b {]}
&=& \ri \ve_{abc} (\g^b)_\b{}^\g \cC^c \cD_\g
+ (\g_a)_\b{}^\g \cS \cD_\g 
- \ri (\g_a)_{\b\g} \bar \cR \bar \cD^\g
+ \ri (\g_a)_\b{}^\g \cD_{(\g} \cC_{\d\r)} M^{\d\r} \non\\
&&- \frac{1}{3} (2 \cD_\b \cS + \ri \bar \cD_\b \bar \cR) M_a
- \frac{2}{3} \ve_{abc} (\g^b)_\b{}^\a (2 \cD_\a \cS + \ri \bar \cD_\a \bar \cR) M^c \non\\
&&+ \frac{\ri}{2} \Big( 
(\g_a)^{\a\g} \cD_{(\a} \cC_{\b\g)}
+ \frac{1}{3} (\g_a)_\b{}^\g (8 \ri \cD_\g \cS - \bar \cD_\g \bar \cR)
\Big) J \ ,
\eea
\esubeq
where the ${\rm U}(1)_{R}$ charges of the torsion superfields  $\cR$, $\bar \cR$ and $\cC_{\a\b}$
are $-2$, $+2$ and 0, respectively.
They also satisfy the Bianchi identities
\bea 
\cD_\a \bar \cR &=& 0 \ , \quad ( \bar \cD^2 - 4\cR) \cS=0\, \quad
\cD^\b \cC_{\a\b} =- \frac{1}{2} (\bar \cD_{\a} \bar \cR + 4 \ri \cD_{\a} \cS) \ ,
\label{BItypeIC}
\eea
Throughout this paper, we make use of the definitions
$\cD^2:=\cD^\a\cD_\a$ and $\cDB^2:=\cDB_\a\cDB^\a$.

The algebra of covariant derivatives given by \eqref{algebra-final}
does not change under the super-Weyl transformation
\cite{KLT-M11, KT-M11}
\bsubeq  \label{super-WeylN=2}
\bea
\cD'_\a&=&\re^{\hf\s}\Big(\cD_\a+\cD^{\g}\s
M_{\g\a}-\cD_{\a} \s J\Big)~,
\\
\cDB'_{\a}&=&\re^{\hf\s}\Big(\cDB_{\a}+\cDB^{\g}\s {M}_{\g\a}
+ \cDB_{\a}\s J\Big)~,
\\
\cD'_{a}
&=&\re^{\s}\Big(
\cD_{a}
-\frac{\ri}{2}(\g_a)^{\g\d}\cD_{\g}\s\cDB_{\d}
-\frac{\ri}{2}(\g_a)^{\g\d}\cDB_{\g}\s\cD_{\d} 
+\ve_{abc}\cD^b \s M^c \non\\
&&~~~~~-\frac{\ri}{2}(\cD^{\g}\s)\cDB_{\g}\s M_{a}
- \frac{\ri}{24} (\g_a)^{\g\d} \re^{- 3 \s} [\cD_\g , \bar \cD_\d] \re^{3 \s}
J
\Big)~,
\eea
which induces the following transformation of the torsion tensors:
\bea 
\cS'&=&\re^{\s}\Big(
\cS
+\frac{\ri}{4}\cD^\g\cDB_{\g}\s
\Big)~,
 \label{2.11d} \\
\cC'_{a}&=&
\Big( \cC_a + \frac{1}{8} (\g_a)^{\g\d} [\cD_\g , \bar \cD_\d] \Big) \re^\s
~,
\\
\cR' &=&
- \frac{1}{4} \re^{2 \s} (\bar \cD^2 - 4 \cR) \re^{- \s}
~,
\label{2.11f}
\eea
\esubeq
where the  parameter $\s$ is an arbitrary real scalar superfield.
The super-Weyl invariance \eqref{super-WeylN=2}
is intrinsic to conformal supergravity. 
For every supergravity-matter system, 
its action is required to be a super-Weyl invariant functional  
of the supergravity multiplet coupled to certain conformal
compensators, see \cite{KLT-M11,KT-M11} for more details.

The $\cN=2$ supersymmetric extension of the Cotton tensor 
\cite{Kuzenko12} is given by 
\bea 
\cW_{\a\b} = - \frac{\ri}{4} [\cD^\g , \bar\cD_\g] \cC_{\a\b}
+ \hf [\cD_{(\a} , \bar\cD_{\b)}] \cS + 2 \cS \cC_{\a\b}  ~.
\label{N=2covsuperspaceCotton}
\eea
It may be checked that $\cW_{\a\b} $
transforms homogeneously,
\bea 
\cW'_{\a\b} = \re^{2 \s} \cW_{\a\b} \ ,
\label{2.9}
\eea
under \eqref{super-WeylN=2}.
The super-Cotton tensor  obeys the Bianchi identities \cite{BKNT-M1}
\bea
 \bar \cD^\b \cW_{\a\b} = \cD^\b \cW_{\a\b} =0~.
\eea
The curved superspace is conformally flat if and only if $\cW_{\a\b} =0$
\cite{BKNT-M1}.


\subsection{Primary superfields} \label{subsection2.2}

Let $T_{\a(n) } := T_{\a_1 \dots \a_n}
=T_{(\a_1 \dots \a_n)}$ be a symmetric rank-$n$ spinor superfield 
of ${\rm U(1)}_R$ charge $q$, 
\bea
J T_{\a(n) }  = q T_{\a(n) } ~.
\eea
The $T_{\a(n) } $
is said to be super-Weyl primary of dimension $d$ if its infinitesimal super-Weyl transformation law is 
\bea
\d_\s T_{\a(n) }  = d \s T_{\a(n) } ~.
\eea
As follows from \eqref{2.9}, the super-Cotton tensor is super-Weyl primary of dimension
$+2$.  We now introduce several types of primary superfields that will be important 
for our subsequent consideration. 

A symmetric rank-$n$ spinor superfield $G_{\a(n)}$ is called longitudinal linear if
it obeys the following first-order constraint
\bea
 \bar \cD_{(\a_1} G_{\a_2 \dots \a_{n+1} )} = 0  ~, \label{2.111}
 \eea
 which implies 
 \bea
 \big(\bar \cD^2+2n\cR \big)G_{\a(n)} &=& 0~. \label{2.12}
 \eea
 If $G_{\a(n)}$ is super-Weyl primary, then the constraint \eqref{2.111} 
 is consistent provided 
 the dimension $d_{G_{(n)} }$ and ${\rm U(1)}_R$ charge $q_{G_{(n)} }$ 
 of $G_{\a(n)}$ are related to each other as follows:
 \bea
 d_{G_{(n)} } =  -\frac{n}{2}  -q_{G_{(n)} }~. \label{2.13}
 \eea
 In the scalar case, $n=0$, the constraint \eqref{2.111} becomes the condition 
 of covariant chirality, $\bar \cD_\a G=0$.  
 The dimension $d_{G }$ and ${\rm U(1)}_R$ charge $q_{G}$ of 
 any primary chiral scalar superfield $G$ are related as
 $ d_{G} +q_{G }=0$, in accordance with  \cite{KLT-M11}.
 
 The longitudinal linear superfields form a ring. Given two such superfields
 $G_{\a(n)}$ and $\tilde G_{\a(m)}$, their product 
 $G_{\a(n+m)} := G_{(\a_1 \dots \a_n} \tilde G_{\a_{n+1} \dots \a_{n+m})}$ 
 is  longitudinal linear. If  $G_{\a(n)}$ and $\tilde G_{\a(m)}$ are super-Weyl primary
 superfields,  their product  $G_{\a(n+m)} $ is also super-Weyl primary.

Given a positive integer $n$, a symmetric rank-$n$ spinor superfield 
$\G_{\a(n)}$ is called transverse linear if it obeys the  first-order constraint
 \bea
 \bar \cD^\b \G_{ \b \a_1 \dots \a_{n - 1} } = 0 ~,  \qquad n \neq 0~,   
 \label{2.144}
 \eea
 which implies 
 \bea
 \big(\bar \cD^2-2(n+2)\cR\big)\G_{\a(n)} = 0~. \label{2.15}
 \eea
If $\G_{\a(n)}$ is super-Weyl primary, then the constraint \eqref{2.144} 
 is consistent provided 
 the dimension $d_{\G_{(n)} }$ and ${\rm U(1)}_R$ charge $q_{\G_{(n)} }$ 
 of $\G_{\a(n)}$ are related to each other as follows:
 \bea
 d_{\G_{(n)} } =  1+ \frac{n}{2}-q_{\G_{(n)} }
 ~. \label{2.16}
 \eea
In the $n=0$ case, the constraint \eqref{2.144} is not defined.
However its corollary \eqref{2.15} is perfectly consistent, 
\bea
 \big(\bar \cD^2- 4\cR\big)\G = 0~, 
\label{2.18}
 \eea
 and defines a covariantly  linear scalar superfield $\G$.
 The dimension $d_{\G}$ and ${\rm U(1)}_R$ charge $q_{\G}$ of 
 any primary linear scalar  $\G$ are related as
 $ d_{\G} +q_{\G}=1$, in accordance with  \cite{KLT-M11}.

In the case of 4D $\cN=1$ AdS supersymmetry, 
longitudinal linear and transverse linear superfields 
were pioneered  by Ivanov and Sorin \cite{IS}
who studied the superfield representations of the AdS isometry group
${\rm OSp}(1|4)$. In the framework of 4D $\cN=1$ conformal supergravity,
primary longitudinal linear and transverse linear supermultiplets were
introduced for the first time by Kugo and Uehara \cite{KU}.
Such superfields were used in \cite{KS94,KSP,KS,KO,BHK}
for the description of off-shell massless gauge theories in four and three dimensions.

The constraints \eqref{2.111} and \eqref{2.144} are solved in terms of 
prepotentials $\J_{\a(n-1)}$ and $\F_{\a(n+1)}$ as follows:
\begin{subequations}
\bea
 G_{\a(n)}&=& \bar \cD_{(\a_1}
{ \J}_{ \a_2 \dots \a_{n}) } ~, 
\label{2.19a}\\
 \G_{\a(n)}&=& \bar \cD^\b
{ \Phi}_{(\b \a_1 \dots \a_n )} ~.
\label{2.19b}
\eea
\end{subequations}
Provided the constraints \eqref{2.111} and \eqref{2.144} are the only conditions
imposed on $ G_{\a(n)}$ and $\G_{\a(n)}$ respectively, 
the prepotentials $\J_{\a(n-1)}$ and $\F_{\a(n+1)}$ can be chosen 
to be  unconstrained complex, and  
are defined modulo
gauge transformations of the form:
\begin{subequations} 
\bea
\d_\z \J_{\a(n-1)} 
&=&  \bar \cD_{(\a_1 }
{ \z}_{\a_2 \dots \a_{n-1})} ~, \label{2.20a}\\
\d_\x \Phi_{\a(n+1)} &=&  \bar \cD^\g
{ \x}_{(\g \a_1 \dots \a_{n+1})} ~,
\eea
\end{subequations}
with the gauge parameters ${\z_{\a(n-2)}}$ and $\x_{\a(n+2)}$
being unconstrained. If the linear superfields $ G_{\a(n)}$
and  $\G_{\a(n)}$ are super-Weyl primary, then their prepotentials 
$\J_{\a(n-1)}$ and $\F_{\a(n+1)}$ can also be chosen to be super-Weyl 
primary. 

Given two linear superfields $ G_{\a(n+1)}$ and  $\G_{\a(n)}$ such that 
 their ${\rm U(1)}_R$ charges are constrained by
$q_{G_{(n+1)} } +q_{\G_{(n)} } =-1$, 
we can define a gauge-invariant and super-Weyl-invariant $BF$ term
\bea
I_{BF}^{(n)} =\int  
\rd^{3|4} z\, E\, \J^{\a(n)} \G_{\a(n)}
= -(-1)^n \int  
\rd^{3|4} z
\, E\,
\F^{\a(n+1)} G_{\a(n+1)}~, 
\eea
where the superspace integration measure is $\rd^{3|4} z:= \rd^3x \rd^2 \q  \rd^2 \bar \q$
and $ E^{-1} := {\rm Ber} (E_A{}^M)$.

In the $n=0$ case, the prepotential solution \eqref{2.19b} 
is still valid. The prepotential $\F_\a$ can be chosen to be unconstrained complex  
provided the constraint \eqref{2.18} is the only condition imposed on $\G$. 
However, if we are dealing with a real linear superfield, 
\bea
 \big(\bar \cD^2- 4\cR\big)L = 0~,  \qquad \bar L =L~,
\label{2.22}
 \eea
then the constraints are solved \cite{KLT-M11} in terms of 
an unconstrained real prepotential $V$,  
\bea
L= \ri {\bar \cD}^\a \cD_\a V~,  \qquad \bar V =V~,
\label{G-prep}
\eea
which is defined modulo gauge transformations of the form:
\bea 
\d V = \l + \bar \l~, \qquad  \cJ \l =0~, \quad {\bar \cD}_\a \l =0~.
\eea
If $L$ is super-Weyl primary, then eq. \eqref{2.16} tells us that
that the dimension of $L$ is $+1$. 
In this case it is consistent to consider the gauge prepotential $V$ 
to be inert under the super-Weyl transformations \cite{KLT-M11}, 
$\d_\s V =0$.

Let us assume that the background curved superspace
allows the existence of 
a real transverse linear superfield 
${\mathfrak W}_{\a(n)} = \bar{\mathfrak W}_{\a(n)} $, 
\bea
\bar \cD^\b {\mathfrak W}_{\b \a_1 \dots \a_{n-1}}=0~, \qquad 
 \cD^\b {\mathfrak W}_{\b \a_1 \dots \a_{n-1}}=0~.
 \label{2.25}
 \eea
 Then it is automatically conserved, 
 \bea
 \cD^{\b\g} {\mathfrak W}_{\b\g\a_1 \dots \a_{n-2}}=0~.
 \label{2.26}
 \eea
 in accordance with \eqref{algebra-final-b}.
The super-Cotton tensor $\cW_{\a\b}$ is an example of such supermultiplets.
If ${\mathfrak W}_{\a(n)}$ is super-Weyl primary, then its dimension is 
equal to $(1+n/2)$, in accordance with \eqref{2.16}.
As will be shown in the next subsection, a solution to \eqref{2.25}
in terms of an unconstrained prepotential exists for every conformally flat superspace.


\subsection{Conformal gauge superfields}\label{subsection2.3}

Let $n$ be a positive integer.
A real symmetric rank-$n$ spinor superfield ${\mathfrak H}_{\a(n) } $ 
is said to be a conformal 
gauge supermultiplet
if (i)  it is super-Weyl primary of dimension $(-{n}/{2})$, 
\bea
\d_\s {\mathfrak H}_{\a(n)} = -\frac{n}{2} \s {\mathfrak H}_{\a(n)}~;
\label{2.27}
\eea
and (ii) it is defined modulo gauge transformations of the form
\bea
\d_\l {\mathfrak H}_{\a(n) } =\bar  \cD_{(\a_1} \l_{\a_2 \dots \a_n) }
-(-1)^n\cD_{(\a_1} \bar \l_{\a_2 \dots \a_n) }~,
\label{2.28}
\eea
with the  gauge parameter $\l_{\a(n-1)}$
being unconstrained complex.
The dimension of ${\mathfrak H}_{\a(n) } $   in \eqref{2.27}
is uniquely fixed by requiring the longitudinal 
linear superfield $g_{\a(n)} =\bar  \cD_{(\a_1} \l_{\a_2 \dots \a_n) }$
 in the right-hand side of \eqref{2.28} to be super-Weyl primary. Indeed,  
the gauge parameter $g_{\a(n)}$ must be neutral with respect to the 
$R$-symmetry  group ${\rm U(1)}_R$ since ${\mathfrak H}_{\a(n) } $ is real,
and then the dimension of  $g_{\a(n)}$ is equal to $(-n/2)$, in accordance with
\eqref{2.13}.

Starting  with ${\mathfrak H}_{\a(n) } $ one can construct its real descendant
${\mathfrak W}_{\a(n)} ( {\mathfrak H}) =\cA {\mathfrak H}_{\a(n)}$, 
where $\cA$ is a linear differential operator involving $\cD_A$,
the torsion superfields and their covariant derivatives,
with the following the properties: 
\begin{enumerate}

\item
${\mathfrak W}_{\a(n)}$ is super-Weyl primary of dimension $(1+n/2)$, 
\bea
\d_\s {\mathfrak W}_{\a(n)} = \big(1+\frac{n}{2} \big)  \s {\mathfrak W}_{\a(n)}~.
\label{2.29}
\eea

\item
The gauge variation of ${\mathfrak W}_{\a(n)}$ vanishes 
if the superspace is conformally flat,
\bea
\d_\l {\mathfrak W}_{\a(n)} = O\big( \cW_{(2)}\big)~,
\eea 
where $\cW_{(2)}$ is the super-Cotton tensor \eqref{N=2covsuperspaceCotton}.

\item
 ${\mathfrak W}_{\a(n)}$ is divergenceless if the superspace is conformally flat,
\bea
\bar \cD^{\b} {\mathfrak W}_{\b \a(n-1)} = O\big( \cW_{(2)}\big)~,\qquad
\cD^{\b} {\mathfrak W}_{\b \a(n-1)} = O\big( \cW_{(2)}\big)~.
\eea 
Here $O\big( \cW_{(2)}\big)$ stands for contributions containing the super-Cotton tensor and
its covariant derivatives.
\end{enumerate}
In general, ${\mathfrak W}_{\a(n)} ( {\mathfrak H}) $ is uniquely defined 
modulo a normalisation and contributions involving the super-Cotton tensor \eqref{N=2covsuperspaceCotton}.
 
Suppose that the background  curved superspace $\cM^{3|4}$
is conformally flat, 
\bea
\cW_{\a \b}=0~.
\label{2.32}
\eea
Then  ${\mathfrak W}_{\a(n)}  ( {\mathfrak H})$ is gauge invariant,
\bea
\d_\l {\mathfrak W}_{\a(n)}&=&0~,
\eea
and obeys the conservation equations \eqref{2.25}.
These properties and the super-Weyl transformation laws
\eqref{2.27} and  \eqref{2.29} imply
that the action\footnote{The super-Weyl transformation of the superspace 
density is $\d_\s E = -\s E$.} 
\bea
{S}_{\rm{SCS}}
[ {\mathfrak H}_{(n)}] 
= - \frac{\ri^n}{2^{\left \lfloor{n/2}\right \rfloor +1}}
   \int \rd^{3|4}z \, E\,
 {\mathfrak H}^{\a(n)} 
{\mathfrak W}_{\a(n) }( {\mathfrak H}) 
\label{2.34}
\eea
is gauge and super-Weyl invariant, 
\bea
\d_\l {S}_{\rm{SCS}}
[{\mathfrak H}_{(n)}] =0~, \qquad 
\d_\s {S}_{\rm{SCS}}
[ {\mathfrak H}_{(n)} ] =0~.
\eea

In accordance with the results of \cite{Kuzenko12,KNT-M},
it is natural to think of ${\mathfrak H}_{\a\b}$ and 
${\mathfrak W}_{\a\b}  ( {\mathfrak H})$ 
as the linearised prepotential 
for $\cN=2$ conformal supergravity  and the linearised super-Cotton tensor
 respectively. It is worth recalling that \eqref{2.32}
is the equation of motion for conformal 
supergravity. The functional \eqref{2.34} 
is proportional to the linearised action for conformal supergravity, 
which is obtained by linearising the nonlinear action for $\cN=2$ conformal supergravity
\cite{RvN,BKNT-M2} around a stationary point defined by \eqref{2.32}.
We can interpret ${\mathfrak W}_{\a(n)}$ to be a linearised higher-spin 
super-Cotton tensor.
We now turn to constructing 
${\mathfrak W}_{\a(n)}$
on a conformally flat superspace. 

In Minkowski superspace, the linearised higher-spin 
super-Cotton tensors were constructed in \cite{KO}, and here we reproduce
these results.
Associated with a real prepotential $H_{\a(n)} = H_{\a_1 \dots \a_n}$ 
is the following real symmetric rank-$n$ spinor descendant
\bea
&&W_{\a (n)}  (H)
= \frac{1}{2^{n-1}} 
\sum\limits_{j=0}^{\left \lfloor{n/2}\right \rfloor}
\bigg\{
\binom{n}{2j} 
\Delta  \Box^{j}\pa_{(\a_{1}}{}^{\b_{1}}
\dots
\pa_{\a_{n-2j}}{}^{\b_{n-2j}}H_{\a_{n-2j+1}\dots\a_{n})\b_1 \dots\b_{n-2j}}~~~~
\nonumber \\
&&\qquad \qquad +
\binom{n}{2j+1}\Delta^{2}\Box^{j}\pa_{(\a_{1}}{}^{\b_{1}}
\dots\pa_{\a_{n-2j -1}}{}^{\b_{n-2j -1}}H_{\a_{n-2j}\dots\a_{n})
\b_1 \dots \b_{n-2j -1} }\bigg\}~,~~~~~
\label{eq:HSFSUniversal}
\eea
where 
\bea
\D = \frac{\ri}{2} D^\a \bar D_\a~,
\label{Delta}
\eea
and $D_A =(\pa_a,D_\a,\DB^\a)$ are the covariant derivatives for Minkowski superspace, 
\bea
&&\pa_a=\frac{\pa}{\pa x^a}~,~~~
D_\a=\frac{\pa}{\pa\q^\a}+\ri\qb^\b(\g^a)_{\a\b}\pa_a
~,~~~
\DB_\a=-\frac{\pa}{\pa\qb^\a}-\ri\q^\b(\g^a)_{\a\b}\pa_a~.
\label{2.38}
\eea
The field strength \eqref{eq:HSFSUniversal}
is invariant, 
\bea
\d_\l W_{\a(n)} =0~,
\eea  
under the gauge transformations 
\bea
\d_\l H_{\a(n)} = 
\bar  D_{(\a_1} \l_{\a_2 \dots \a_n) }
-(-1)^n D_{(\a_1} \bar \l_{\a_2 \dots \a_n) }~,
\eea
where the gauge parameter  $\l_{\a(n-1)}  $ is unconstrained complex. 
The field strength  \eqref{eq:HSFSUniversal} is conserved, 
\bea
D^\b W_{\b \a_1 \dots \a_{n-1} } = \bar D^\b W_{\b \a_1 \dots \a_{n-1} } = 0~.
\eea
Making use of $W_{\a(n)}$ allows us to construct 
the  higher-spin 
super-Cotton tensor ${\mathfrak W}_{\a(n)}$ in any 
conformally flat superspace $\cM^{3|4}$.

In accordance with \eqref{super-WeylN=2}, for a
conformally flat superspace $\cM^{3|4}$ we can choose a local frame in which  
the covariant derivatives have the form
\bsubeq \label{2.42}
\bea
\cD_\a&=&\re^{\hf\s}\Big(D_\a+D^{\g}\s M_{\g\a}-D_{\a }\s  J\Big)~,
\\
\bar \cD_{\a}&=&\re^{\hf\s}\Big(\DB_{\a}+\DB^{\g}\s M_{\g\a}
+\DB_{\a}\s J \Big)~,
\\
\cD_{a}
&=&\re^{\s}\Big(
\pa_{a}
-\frac{\ri}{2}(\g_a)^{\g\d} D_{(\g}\s \DB_{\d)}
-\frac{\ri}{2}(\g_a)^{\g\d} \DB_{(\g}\s D_{\d)}
+\ve_{abc}\pa^b\s  M^c
\non\\
&&
+\frac{\ri}{2}(D_{\g}\s)\DB^{\g}\s M_{a}
- \frac{\ri}{24} (\g_a)^{\g\d} \re^{- 3 \s} [D_\g , \bar D_\d] \re^{3 \s} J
\Big)
~,
\label{Da_2-0}
\eea
\esubeq
for some real scale factor $\s$.
Then, in accordance with \eqref{2.29},
the  higher-spin 
super-Cotton tensor ${\mathfrak W}_{\a(n)}$ in $\cM^{3|4}$
is related to the flat-space one, eq. \eqref{eq:HSFSUniversal},
by the rule
\bea
{\mathfrak W}_{\a(n)} = \re^{(1+\frac{n}{2})\s} W_{\a(n)}~. 
\label{2.43}
\eea
Similarly, eq. \eqref{2.27} tells us that the prepotentials 
${\mathfrak H}_{\a(n)}$ and ${H}_{\a(n)}$ 
can be chosen to be related  to each other by 
\bea
{\mathfrak H}_{\a(n)} = \re^{-\frac{n}{2}\s} H_{\a(n)}~. 
\eea
In general, it is a difficult technical problem to express ${\mathfrak W}_{\a(n)}$ 
in terms of the covariant derivatives $\cD_A$ and the gauge prepotential  
${\mathfrak H}_{\a(n)} $, for arbitrary $n$.

There exists a refined version of the representation \eqref{2.42} 
for those conformally flat superspaces which are characterised by 
the condition
\bea
\cS=0~. \label{2.45}
\eea
This family includes the (1,1) AdS superspace defined by the 
(anti)commutation relations \eqref{(1,1)AdS}.
If \eqref{2.45} holds, then eq. \eqref{2.11d} tells that 
the scale factor in \eqref{2.42} is constrained,
\bea
D^\g \bar D_\g \s=0 \quad \implies \quad \s = \eta +\bar \eta~, \qquad
\bar D_\a \eta=0~,
\eea
with the chiral scalar $\eta$ being, in principle, arbitrary. 
Now, applying a local $R$-symmetry transformation 
\bea
\cD_A ~\to ~ {\bm \cD}_A = \re^{-( \eta - \bar\eta)J } \cD_A \re^{(\eta- \bar \eta)J}
\eea
leads to covariant derivatives without U(1)${}_R$ connection.
The resulting covariant derivatives are
\bsubeq \label{sW+U(1)}
\bea
{\bm \cD}_\a&=&
\re^{\frac{1}{2}(3\bar{\eta}-\eta)}
\Big(D_\a+D^{\g} \eta M_{\g\a}\Big)~,
\\
\bar {\bm \cD}_\a&=&
\re^{\frac{1}{2}(3\eta-\bar{\eta})}
\Big(\bar D_\a+\bar D^{\g}\bar{\eta} M_{\g\a}\Big)~,
\\
{\bm \cD}_{a}&=&\re^{\eta+\bar{\eta}}\Big(
\pa_{a}
-\frac{\ri}{2}(\g_a)^{\a\b}D_{\a}\eta \bar D_{\b}
-\frac{\ri}{2}(\g_a)^{\a\b}\bar D_{\a}\bar{\eta} D_{\b}
\non\\
&&~~~~~~~
+\ve_{abc}\pa^b(\eta+\bar{\eta}) M^c
+\frac{\ri}{2}(D_{\g}\eta)(\bar D^{\g}\bar{\eta}) M_{a}
\Big)
~.
\eea
\esubeq
In the case of (1,1) AdS superspace, the scale factor $\eta$ 
was computed in \cite{KT-M11}.


\section{(1,1) AdS superspace} \label{section2}

In this section we give a brief summary of the key results
concerning the (1,1) AdS superspace \cite{KT-M11}, 
as well as elaborate on superfield representations of the (1,1) AdS 
isometry group.
The covariant derivatives of ${\rm AdS}^{(3|1,1)}$ satisfy the following algebra \cite{KT-M11}:
\begin{subequations}  \label{1.2}
\bea
&& \qquad \{ \cD_\a , \bar \cD_\b \} = -2\rm i \cD_{\a \b} ~, \\
&& \qquad \{\cD_\a, \cD_\b \} = -4\bar \m\, M_{\a \b}~, \qquad
\{ {\bar \cD}_\a, {\bar \cD}_\b \} = 4\m\,M_{\a \b}~, \\
&& \qquad [ \cD_{ \a \b }, \cD_\g ] = -2 \rm i \bar \m\,\ve_{\g (\a} \bar \cD_{\b)}~,  \qquad
\,\,[\cD_{ \a \b }, { \bar \cD}_{\g} ] = 2 \rm i \m\,\ve_{\g (\a} \cD_{\b)}~,   \\
&&\quad \,\,\,\,[ \cD_{\a \b} , \cD_{ \g \d } ] = 4 \bar \m \m \Big(\ve_{\g (\a} M_{\b) \d}+ \ve_{\d (\a} M_{\b) \g}\Big)~,  
\eea
\end{subequations} 
with $\m\neq 0$ being a  complex parameter.
As compared with \eqref{(1,1)AdS}, we have denoted $\cR = \m$.
This notation will be used in the remainder of this paper.

The covariantly  transverse linear and longitudinal linear superfields
on an arbitrary  supergravity background were described in 
the previous section. In the case of (1,1) AdS superspace, 
such superfields play an important role.
One can define projectors $P^{\perp}_{n}$ and $P^{||}_{n}$
on the spaces of  transverse linear and longitudinal linear superfields, respectively.
The projectors are 
\begin{subequations}
\bea
P^{\perp}_{n}&=& \frac{1}{4 (n+1)\m} (\bar \cD^2+2n\m) ~,\\
P^{||}_{n}&=&- \frac{1}{4 (n+1)\m} (\bar \cD^2-2(n+2)\m ) ~,
\eea
\end{subequations} 
with the properties 
\bea
\big(P^{\perp}_{n}\big)^2 =P^{\perp}_{n} ~, \quad 
\big(P^{||}_{n}\big)^2=P^{||}_{n}~,
\quad P^{\perp}_{n} P^{||}_{n}=P^{||}_{n}P^{\perp}_{n}=0~,
\quad P^{\perp}_{n} +P^{||}_{n} ={\mathbbm 1}~.
\eea

Given a complex tensor superfield $V_{\a(n)} $ with $n \neq 0$, 
it can be represented
as a sum of transverse linear and longitudinal linear multiplets, 
\bea
V_{ \a(n)} = &-& 
\frac{1}{2 \mu (n+2)} \cDB^\g \cDB_{(\g} V_{ \a_1 \dots  \a_n)} 
- \frac{1}{2 \mu (n+1)} \cDB_{(\a_1} \cDB^{|\g|} V_{ \a_2 \dots \a_{n} ) \g} 
~ . ~~~
\eea
Choosing $V_{ \a(n)} $ to be  
longitudinal linear ($G_{ \a(n)} $)
or transverse linear ($\G_{ \a(n)} $), the above identity 
gives the relations \eqref{2.19a} and \eqref{2.19b}
for some prepotentials $\J_{\a(n-1)}$ and $\F_{\a(n+1)} $, respectively.

In accordance with the general formalism of \cite{BK},
the isometries of ${\rm AdS}^{(3|1,1)} $ are generated by 
those real supervector fields $\l^A E_A $ which obey 
the Killing equation 
\bea
\Big{[}\L+\hf l^{ab}M_{ab},\cD_C\Big{]}=0~,
\eea
where 
\bea
\L= \l^A \cD_A =\l^a\cD_a+\l^\a\cD_\a+\bar \l_\a\bar \cD^\a~,
\qquad \overline{\l^a}=\l^a
\eea 
and $l^{ab}$ is some local Lorentz parameter.
As demonstrated in \cite{KT-M11},
this equation implies that the parameters $\l^\a$ and $l^{ab}$ 
are uniquely expressed in terms of the vector  $\l^a$,
\bea
\l_\a =\frac{\ri}{6} \bar \cD^\b \l_{\a\b}~,\qquad
l_{\a\b} =2\cD_{(\a}\l_{\b)}~,
\label{2.5}
\eea
and the vector parameter obeys the equation
\bea
\cD_{(\a}\l_{\b\g)}=0 \quad \Longleftrightarrow \quad 
\bar\cD_{(\a}\l_{\b\g)}=0~.
\eea
In comparison with the 3D $\cN=2$ Minkowski superspace,
the specific feature of ${\rm AdS}^{(3|1,1)} $ is that any two of the three parameters 
 $\{ \l_{\ab}, \l_\a, l_{\a\b}\}$ are expressed in terms of the third parameter, 
in particular
\bea
\l_{\a\b} =\frac{\ri}{\m} \bar \cD_{(\a}\l_{\b)}~,\qquad 
\l_\a =
-\frac{1}{12 \bar \m} \cD^\b l_{\a\b}
~.
\label{2.7}
\eea
From \eqref{2.5} and \eqref{2.7} we deduce
\bea
\bar \cD^\a\l_\a=
\cD_\a \l^\a=0~.
\label{1,1-SK_1}
\eea
Every solution $\l^A$ of the above relations is called is a  Killing
supervector field  of ${\rm AdS}^{(3|1,1)} $.
 These supervector fields can be shown to generate  the isometry group of ${\rm AdS}^{(3|1,1)}$,
${\rm OSp(1|2; {\mathbb{R}})} \times {\rm OSp(1|2; {\mathbb{R}})}$,

In Minkowski superspace ${\mathbb M}^{3|4}$, there are two ways to generate supersymmetric invariants, 
one of which corresponds to the 
integration over the full superspace and the other over its chiral subspace. 
In (1,1) AdS superspace, every chiral integral can always be recast as 
a full superspace integral.
Associated with a scalar superfield $\cL$ is the following 
supersymmetric  invariant
\bea
\int \rd^3x \rd^2 \q  \rd^2 \bar \q
\,E\,{\cal L} &=& 
-\frac{1}{4} \int
\rd^3x \rd^2 \q  
\, \cE\, {({\bar \cD}^2 - 4 \m)} {\cal L} ~, \qquad
E^{-1}= {\rm Ber}\, (E_{\rm A}{}^M)~,
\label{3.11}
\eea
where 
$\cE$ denotes the chiral integration measure.
Let $\cL_{\rm c}$ be a covariantly chiral scalar Lagrangian, 
$\bar \cD_\a \cL_{\rm c} =0$. 
It generates a supersymmetric invariant of the form
$
\int \rd^3x \rd^2 \q  \, \cE \,{\cal L}_{\rm c}. 
$
The specific feature
of (1,1) AdS superspace is that the chiral action can equivalently
be written as an integral over the full superspace \cite{KT-M11}
\bea
\int \rd^3x \rd^2 \q  \, \cE \,{\cal L}_{\rm c} 
= \frac{1}{\m} \int \rd^3x \rd^2 \q  \rd^2 \bar \q
\,{E}\, {\cal L}_{\rm c} ~.
\eea
Unlike the flat superspace case, the integral on the right does not vanish in AdS.

Supersymmetric invariant \eqref{3.11} can be reduced to component fields by the rule
\cite{KLRST-M}
\bea
\int \rd^3x \rd^2 \q  \rd^2 \bar \q
\,E\,{\cal L} 
=\frac{1}{16}
\int\rd^3x\,e\,
(\cD^2- 16 \bar \m) ({\bar \cD}^2 - 4 \m) \cL 
\big|~,
\label{comp-ac-1}
\eea
with $e^{-1} := \det (e_a{}^m)$. Here $e_a{}^m$ is the inverse vielbein, which determines 
the torsion-free covariant derivative of AdS space
\bea
\nabla_a=e_a+\hf\o_a{}^{bc} (e)M_{bc}~, \qquad
e_a:=e_a{}^m \pa_m~.
\label{stcd-2}
\eea
In general, the $\q ,\bar \q$-independent component,  $T |_{\q=\bar \q=0}$, 
of a superfield $T(x,\q, \bar \q)$  is denoted $T|$. To complete the formalism 
of component reduction, we only need the following relation 
\bea
\big(\cD_a T \big) \big| = \nabla_a T|~.
\eea

In what follows, we will work with full superspace integrals only and 
make use of the notation $\rd^{3|4} z:= \rd^3x \rd^2 \q  \rd^2 \bar \q$.


\section{Massless half-integer superspin gauge theories in (1,1) AdS superspace} 

The superconformal higher-spin action \eqref{2.34}
 in a conformally flat superspace is formulated in terms of 
 the conformal gauge superfields ${\mathfrak H}_{\a(n)}$. 
The same gauge superfield, at least for $n=2s$, with $s=1,2,\dots,$
can be used to construct massless actions
 in two of the three $\cN=2$ maximally symmetric backgrounds, which are  
 Minkowski superspace and (1,1) AdS superspace.
  Such actions, however, involve
  not only ${\mathfrak H}_{\a(n)}$ but also some compensators.

In Minkowski space, 
there are two off-shell formulations for the massless ${\cal N}= 2$ multiplet 
of half-integer superspin $(s+1/2)$, with $s=2, 3, \ldots$, which are dual to each other \cite{KO}. They are referred to as transverse and longitudinal. Here we extend these gauge theories to (1,1) AdS superspace. 


\subsection{Transverse formulation} \label{subsection4.1}

The transverse formulation 
for the massless superspin-$(s+\hf)$ multiplet 
is realised in terms of the following dynamical variables: 
\be
\cV^\bot_{(s+\hf )} = 
\Big\{ {\mathfrak H}_{\a(2s)}, \G_{\a(2s-2)}, \bar{\G}_{\a(2s-2)} \Big\} ~.
\label{4.1tr}
\ee
Here ${\mathfrak H}_{\a(2s)} = {\mathfrak H}_{(\a_1 \dots \a_{2s})}$ 
is an unconstrained real superfield, and the complex superfield 
$\G_{\a(2s-2)} = \G_{(\a_1 \dots \a_{2s-2})}$
is transverse linear, eq. \eqref{2.144}.
In accordance with \eqref{2.19b},
the constraint on $\G_{\a(2s-2)} $ is
 solved in terms of an unconstrained prepotential $\Phi_{\a(2s-1)}$,
\bea
 \G_{\a(2s-2)}= \bar \cD^\b
{ \Phi}_{(\b \a_1 \dots \a_{2s-2} )} ~,
\label{4.22}
\eea
which is defined modulo gauge transformations of the form 
\be
\d_\x \Phi_{\a(2s-1)} 
=  \bar \cD^\b 
{ \x}_{(\b \a_1 \dots \a_{2s-1})} ~,
\label{tr-prep-gauge}
\ee
with the gauge parameter ${\x_{\a(2s)}}$ being unconstrained.

The dynamical superfields $ {\mathfrak H}_{\a(2s)}$ and $ \G_{\a(2s-2)}$
are postulated to be  defined modulo gauge transformations of the form 
\begin{subequations} \label{tr-gauge-half}
\bea
\d_\l {\mathfrak H}_{\a(2s)}&=& 
{\bar \cD}_{(\a_1} \l_{\a_2 \dots \a_{2s})}-{\cD}_{(\a_1}\bar \l_{\a_2 \dots \a_{2s})}
\equiv
g_{\a(2s)}+\bar{g}_{\a(2s)} ~, \label{H-gauge} \\ 
\d_\l \Gamma_{\a(2s-2)}&=&
-\frac{1}{4}\bar{\cD}^{\b} 
\big( {\cD}^2  +2(2s-1) \bar \m \big)\bar{\l}_{\b\a(2s-2)}
= \frac{s}{2s+1}\bar{\cD}^{\b}\cD^{\g}\bar g_{(\b \g \a_1 \dots \a_{2s-2})}
~,~~~
\label{gamma-gauge}
\eea
\end{subequations}
where the complex gauge parameter $\l_{\a(2s-1)}$ is unconstrained. 
The gauge transformation of ${\mathfrak H}_{\a(2s)}$
coincides with \eqref{2.28} for $n=2s$.
From $\d_\l \Gamma_{\a(2s-2)}$ we read off 
the gauge transformation of the prepotential $\Phi_{\a(2s-1)}$
defined by eq. \eqref{4.22}, which  is 
\bea
\d_\l \Phi_{\a(2s-1)} = -\frac{1}{4}
\big( {\cD}^2  +2(2s-1) \bar \m \big)\bar{\l}_{\a(2s-1)}~. \label{eee1}
\eea
Modulo an overall  normalisation factor, there is a unique quadratic action 
which is invariant under 
the gauge transformations \eqref{tr-gauge-half}. It is given by 
\bea
S^{\perp}_{(s+\hf)}
&=& \Big(-\hf \Big)^s \int 
 \rd^{3|4}z \, E
\bigg\{ \frac{1}{8} {\mathfrak H}^{\a(2s)}  \cD^\b ({\bar \cD}^2- 6\mu) \cD_\b 
{\mathfrak H}_{\a(2s)}
\non \\
&&+ 2s(s-1)\bar \m \m {\mathfrak H}^{\a(2s)} {\mathfrak H}_{\a(2s)}
+ {\mathfrak H}^{\a(2s)}\big(\cD_{\a_1} {\bar \cD}_{\a_2} \G_{\a_3 \dots \a_{2s}} 
- {\bar \cD}_ {\a_1} \cD_{\a_2} {\bar \G}_{\a_3 \dots \a_{2s}} \big)
\non \\
&&+ \frac{2s-1}{s} \bar \G^{\a(2s-2)} \G_{\a(2s-2)} 
+ \frac{2s+1}{2s} 
\big(\G^{\a(2s-2)}\G_{\a(2s-2)} + \bar \G^{\a(2s-2)} \bar \G_{\a(2s-2)}\big) \bigg\}
\label{tr-action-half}~.~~~~~
\eea
In the flat superspace limit, this action 
reduces to the one derived in \cite{KO}.

The $s=1$ choice was excluded from the above consideration, since 
the constraint  \eqref{2.144} is not defined for $n=0$.
However, as discussed in section 2, the corollary \eqref{2.15}
of  \eqref{2.144} is perfectly consistent for $n=0$ and defines  a covariantly 
transverse linear scalar superfield \eqref{2.18}, 
\bea
(\bar \cD^2 -\m ) \G=0~.
\label{4.66}
\eea
We therefore postulate  $\G$ and its conjugate $\bar \G$
to be the compensators in the $s=1$ case. 
Choosing $s=1$ in the gauge transformation law \eqref{tr-gauge-half} 
gives
\begin{subequations}
\bea
\d_\l {\mathfrak H}_{\a \b}&=& 
{\bar \cD}_{(\a} \l_{\b)}-{\cD}_{(\a}\bar \l_{\b)} ~, 
\\ 
\d_\l \Gamma&=&
-\frac{1}{4}\bar{\cD}^{\b} 
\big( {\cD}^2  +2 \bar \m \big)\bar{\l}_{\b}~.
\eea
\end{subequations}
The variation $\d_\l \G$ is compatible with 
the constraint \eqref{4.66}, that is $(\bar \cD^2 -\m ) \d_\l \G=0$.
Finally, choosing $s=1$ in \eqref{tr-action-half} gives the linearised 
action for non-minimal (1,1) AdS supergravity, 
which was originally derived in section 9.2 of \cite{KT-M11}.


\subsection{Longitudinal formulation}  \label{subsection4.2}

The longitudinal formulation 
for the massless superspin-$(s+\hf)$ multiplet 
is described in terms of the following variables:
\bea
\cV^{\|}_{(s+\hf)} = \Big\{ {\mathfrak H}_{\a(2s)}, G_{\a(2s-2)}, \bar{G}_{\a(2s-2)} \Big\} ~.
\eea
Here ${\mathfrak H}_{\a(2s)}$ is the same as in \eqref{4.1tr},
and the complex superfield 
$G_{\a(2s-2)}$
is longitudinal linear, eq.  \eqref{2.111}. 
In accordance with \eqref{2.19a}, 
the constraint \eqref{2.111} can be solved in terms of 
an unconstrained complex prepotential $\J_{\a(2s-3)}$,
\be
 G_{\a(2s-2)}= \bar \cD_{(\a_1}
{ \J}_{ \a_2 \dots \a_{2s-2}) } ~,
\label{long-prep}
\ee
which is defined modulo gauge transformations of the form 
\be
\d_\z \J_{\a(2s-3)} 
=  \bar \cD_{(\a_1 }
{ \z}_{\a_2 \dots \a_{2s-3})} ~,
\label{long-prep-gauge}
\ee
with the gauge parameter ${\z_{\a(2s-4)}}$ being unconstrained complex.

The longitudinal formulation may be obtained from the transverse one, 
developed in the previous subsection, 
by performing a superfield duality transformation. 
Starting from 
the action $S^{\perp}_{(s+\hf)}=S^{\perp}_{(s+\hf)}[{\mathfrak H},\Gamma, \bar \G]$, 
eq. \eqref{tr-action-half}, we introduce a first-order model 
described by the action
\bea
&&S[{\mathfrak H},V, \bar V, G, \bar G] = \Big(-\hf \Big)^s 
\int 
\rd^{3|4}z
\, E \,
\bigg\{\frac{1}{8}{\mathfrak H}^{\a(2s)}{\cD}^{\b}({\bar \cD}^{2}-6\mu){\cD}_{\b}
{\mathfrak H}_{\a(2s)}
\non \\
 &&\qquad 
+ 2s(s-1)\mu\bar{\mu} {\mathfrak H}^{\a (2s)}{\mathfrak H}_{\a (2s)} 
+ {\mathfrak H}^{\a(2s)}\Big(\cD_{\a_{1}}\bar{\cD}_{\a_{2}}V_{\a(2s-2)}
-\bar{\cD}_{\a_{1}}\cD_{\a_{2}}\bar{V}_{\a(2s-2)}\Big)\non \\
&&\qquad + \frac{2s-1}{s} \bar V^{\a(2s-2)} V_{\a(2s-2)}
+\frac{2s+1}{2s}\Big(V^{\a(2s-2)} V_{\a(2s-2)}+\bar{V}^{\a(2s-2)}\bar V_{\a(2s-2)}\Big) 
 \non \\
&& \qquad -\frac{2}{s}\Big(G^{\a(2s-2)} V_{\a(2s-2)} 
+ \bar{G}^{\a(2s-2)} \bar{V}_{\a(2s-2)}\Big)\bigg\}~.
\label{1storder-half}
\eea
Here $V_{\a(2s-2)}$ is an unconstrained complex superfield, 
and $G_{\a(2s-2)}$ is given by \eqref{long-prep}.
The first-order action is invariant under the gauge transformation 
\eqref{H-gauge} accompanied with 
\begin{subequations}
\bea
\delta_\l V_{\a(2s-2)} &=&\delta_\l \G_{\a(2s-2)}~, \\
\delta_\l G_{\a(2s-2)}
&=& -\frac{1}{4}\big( \bar{\cD}^{2} -4s\m\big)\cD^{\b}\l_{\a(2s-2)\b}
+\ri (s-1)\bar{\cD}_{(\a_{1}} \cD^{|\b\g|} \l_{\a_2 \dots \a_{2s-2})\b\g} \non \\
&=& s\Big( \frac{1}{2s+1}\cD^{\b}\bar{\cD}^{\g}
+ \ri \cD^{\b\g} \Big) g_{\b\g\a(2s-2)}~,
\label{G-gauge}
\eea
\end{subequations}
where $\delta_\l \G_{\a(2s-2)}$ is given by \eqref{gamma-gauge}.
From \eqref{G-gauge} we  read off the transformation law of 
the prepotential $\J_{\a(2s-2)}$, eq. \eqref{long-prep}, which is
\bea
\delta_\l \J_{\a(2s-3)}&=&-\hf \left({\bar \cD}^\b \cD^\g -2 \ri (s-1) \cD^{\b \g} \right)
 \l_{ \b \g \a(2s-2)}~.
\label{prep-gauge-L}
\eea

Varying the action \eqref{1storder-half} with respect to $\J_{\a(2s-2)}$
implies that  
$V_{\a(2s-2)} =\G_{\a(2s-2)}$, and then $S[{\mathfrak H},V, \bar V, G, \bar G] $
reduces to the transverse action \eqref{tr-action-half}.
This means that the theories \eqref{tr-action-half}
and \eqref{1storder-half} are equivalent.
On the other hand, $V_{\a(2s-2)}$ and its conjugate $\bar V_{\a(2s-2)}$
are auxiliary since they 
appear in the action \eqref{1storder-half} without derivatives. 
Integrating out these auxiliary superfields 
 leads to the following dual theory 
\bea
S^{\|}_{(s+\hf)}
&=& \Big(-\hf \Big)^{s}\int 
\rd^{3|4}z
\, E \bigg\{\frac{1}{8}{\mathfrak H}^{\a(2s)}
\cD^{\b}(\bar{\cD}^{2}-6\mu)\cD_{\b}
{\mathfrak H}_{\a(2s)} \non \\
&& +2s(s-1)\mu\bar{\mu} {\mathfrak H}^{\a(2s)}{\mathfrak H}_{\a(2s)}
-\frac{1}{16}([\cD_{\b},\bar{\cD}_{\g}]{\mathfrak H}^{\b \g \a(2s-2)})
[\cD^{\d},\bar{\cD}^{\r}]{\mathfrak H}_{\d \r \a(2s-2)}
 \non \\
&& +\frac{s}{2}(\cD_{\b \g}{\mathfrak H}^{\b \g \a(2s-2)})
\cD^{\d \r}{\mathfrak H}_{\d \r \a(2s-2)}
\non \\
&&+  \frac{2s-1}{2s} \Big[ \ri
(\cD_{\b \g}{\mathfrak H}^{\b \g \a(2s-2)}) 
\left( G_{\a(2s-2)}-\bar{G}_{\a(2s-2)} \right)
+\frac{1}{s}
\bar{G}^{\a(2s-2)} G_{\a(2s-2)} \Big]
\non \\
&& 
-\frac{2s+1}{4s^{2}}\left(G^{\a(2s-2)}G_{\a(2s-2)}+\bar{G}^{\a(2s-2)} \bar G_{\a(2s-2)}\right)
\bigg\}~.
\label{long-action-half}
\eea
This action is invariant under the gauge transformations 
\begin{subequations} \label{4.166}
\bea
\d_\l {\mathfrak H}_{\a(2s)}&=& 
{\bar \cD}_{(\a_1} \l_{\a_2 \dots \a_{2s})}-{\cD}_{(\a_1}\bar \l_{\a_2 \dots \a_{2s})} ~,\\
\delta_\l G_{\a(2s-2)}&=& 
-\frac{1}{4}\big( \bar{\cD}^{2} -4s\m\big)\cD^{\b}\l_{\a(2s-2)\b}
+\ri (s-1)\bar{\cD}_{(\a_{1}} \cD^{|\b\g|} \l_{\a_2 \dots \a_{2s-2})\b\g} ~.
\eea
\end{subequations}
In the flat superspace limit, this action 
reduces to the one derived in \cite{KO}.

In the $s=1$ case, the compensator $G$ becomes covariantly chiral, $\bar \cD_\a G=0$.
Choosing $s=1$ in \eqref{long-action-half} gives the linearised 
action for minimal (1,1) AdS supergravity, 
which was originally derived in section 9.1 of \cite{KT-M11}, 
provided we identify $G=3\s$. 
Choosing $s=1$ in the gauge transformation law \eqref{4.166} 
gives
\begin{subequations}
\bea
\d_\l {\mathfrak H}_{\a \b}&=& 
{\bar \cD}_{(\a} \l_{\b)}-{\cD}_{(\a}\bar \l_{\b)} ~, \\
\delta_\l G&=& 
-\frac{1}{4}\big( \bar{\cD}^{2} -4\m\big)\cD^{\b}\l_{\b}~.
\eea
\end{subequations}
It is clear that the variation $\delta_\l G$ is covariantly chiral.


\section{Massless integer superspin gauge theories in (1,1) AdS superspace} 

When attempting to develop a Lagrangian formulation for a massless multiplet 
of superspin $s$, where $s=1,2,\dots$, a naive expectation is that 
the dynamical variables of such a theory should consist of 
a conformal gauge superfield ${\mathfrak H}_{\a(2s-1)} =\bar {\mathfrak H}_{\a(2s-1)} $,
introduced in subsection \ref{subsection2.3}, in conjunction with some compensator(s).
Instead, our approach in this section will be based on developing
3D $\cN=2$ analogues of the two dually equivalent off-shell
formulations, the so-called longitudinal and transverse  ones, 
for the  massless ${\cal N}=1$ multiplets of 
integer superspin in AdS${}_4$ \cite{KS94}. Then we will provide a reformulation 
of the longitudinal formulation derived in the next subsection 
in a way similar to the one proposed in the 4D $\cN=1$ AdS case \cite{BHK}.
Such a reformulation naturally leads to the appearance of 
a conformal gauge superfield ${\mathfrak H}_{\a(2s-1)}  $.

\subsection{Longitudinal formulation} \label{subsection5.1}

Given an integer $s \geq 1$, the longitudinal formulation 
for the massless superspin-$s$ multiplet 
is realised in terms of the following dynamical variables:
\bea
\cV^{\|}_{(s)} = \Big\{U_{\a(2s-2)}, G_{\a(2s)}, \bar{G}_{\a(2s)} \Big\} ~.
\label{5.1}
\eea
Here, $U_{\a(2s-2)}$ is an unconstrained real superfield, and the complex superfield $G_{\a(2s)}$ is longitudinal linear, eq.  \eqref{2.111}. 
In accordance with \eqref{2.19a}, 
the constraint \eqref{2.111} can be solved in terms of an  unconstrained complex prepotential $\J_{\a(2s-1)}$,
\bea
G_{\a_1 \dots \a_{2s}} := 
\bar \cD_{(\a_1} \J_{\a_2 \dots \a_{2s})}~,
\label{g2.5}
\eea
which is defined modulo gauge transformations of the form 
\be
\d_\z \J_{\a(2s-1)} =  \bar \cD_{(\a_1 }
{ \z}_{\a_2 \dots \a_{2s-1})} ~,
\ee
with the gauge parameter ${\z_{\a(2s-2)}}$ being unconstrained complex.

We postulate the dynamical superfields $ U_{\a(2s-2)}$ and $ \G_{\a(2s)}$
 to be  defined modulo gauge transformations of the form 
 \begin{subequations}\label{5.4ab}
\bea
\d_L U_{\a(2s-2)}
&=&\cD^{\b}L_{\b \a_1 \dots \a_{2s-2}}-\bar{\cD}^{\b}\bar{L}_{\b \a_1 \dots \a_{2s-2}} 
\equiv \bar \g_{\a(2s-2)}+{\g}_{\a(2s-2)}
~, \label{H-gauge-int} \\
\d_L G_{\a(2s)} 
&=&-\frac{1}{2}\bar{\cD}_{(\a_{1}}\Big(\cD^{2}
-2(2s+1) \bar \m \Big)
L_{\a_2 \dots \a_{2s})}
= \bar{\cD}_{(\a_{1}} \cD_{\a_{2}}\bar \g_{\a_3 \dots \a_{2s})}
~.
\label{int-long-gauge}
\eea
\end{subequations} 
Here the gauge parameter $L_{\a(2s-1)}$ is an unconstrained complex superfield, 
and ${\g}^{\a(2s-2)}:= {\bar \cD}_{\b}\bar L^{\b \a(2s-2)}$ is transverse linear. 
From \eqref{int-long-gauge} we read off the gauge transformation law of the prepotential, 
\bea
\d_L \J_{\a(2s-1)} =-\frac{1}{2} \Big(\cD^{2}
-2(2s+1) \bar \m \Big) L_{ \a (2s-1)}
=\cD_{(\a_1} \cD^{|\b |} L_{\a_2 \dots \a_{2s-1}) \b}~. 
\eea

Modulo an overall  normalisation factor, there is a unique quadratic action 
which is invariant under 
the gauge transformations \eqref{5.4ab}. The action is
\bea
S_{(s)}^{\|}
&=& \Big(-\hf\Big)^{s}\int 
\rd^{3|4}z\, E \,
\bigg\{\frac{1}{8}U^{\a(2s-2)}\cD^{\g}({\bar \cD}^{2}-6\mu)\cD_{\g}U_{\a(2s-2)}
\non \\
&&+\frac{s}{2s+1}U^{\a(2s-2)}\Big(\cD^{\b} {\bar \cD}^{\g}
G_{\b \g \a(2s-2) }
-{\bar \cD}^{\b}{\cD}^{\g}\bar{G}_{\b \g \a(2s-2)} \Big) \non \\
&&+\frac{s}{2s-1}  \bar{G}^{\a(2s)} G_{\a(2s)}
+ \frac{s}{2(2s+1)}\Big(G^{\a(2s)}G_{\a(2s)}+\bar{G}^{\a(2s)}\bar G_{\a(2s)}\Big) 
\non \\
&&+2s(s+1)\mu\bar{\mu} U^{\a(2s-2)}U_{\a(2s-2)}
\bigg\}~.
\label{long-action-int}
\eea
The special $s=1$ case, which corresponds to the massless gravitino  
multiplet, will be studied in more detail in subsection \ref{subsection5.4}.


\subsection{Transverse formulation}

The transverse formulation for the massless superspin-$s$ multiplet 
is realised in terms of the following dynamical variables:
\bea
\cV^{\perp}_{(s)} = \Big\{U_{\a(2s-2)}, \G_{\a(2s)}, \bar{\G}_{\a(2s)} \Big\} ~.
\eea
Here, $U_{\a(2s-2)}$ is the same as in \eqref{5.1}, 
and the complex superfield $\G_{\a(2s)}$ is transverse linear, eq. \eqref{2.144}.
In accordance with \eqref{2.19b},
the constraint on $\G_{\a(2s)} $ is
 solved in terms of an unconstrained prepotential $\Phi_{\a(2s+1)}$,
\bea
 \G_{\a(2s)}= \bar \cD^\b
{ \Phi}_{(\b \a_1 \dots \a_{2s} )} ~,
\label{5.8}
\eea
which is defined modulo gauge transformations of the form 
\bea
\d_\x \Phi_{\a(2s+1)} 
=  \bar \cD^\b 
{ \x}_{(\b \a_1 \dots \a_{2s+1})} ~,
\eea
with the gauge parameter ${\x_{\a(2s+2)}}$ being unconstrained.

The transverse formulation for the massless superspin-$s$ multiplet 
is obtained from the longitudinal one developed in the previous subsection 
by performing a superfield duality transformation.
The first step is to replace the gauge-invariant action \eqref{long-action-int}
with 
the following first-order action
\bea
S_{s}[U,V, \bar V, \G, \bar \G] &=& \Big(-\hf\Big)^{s}\int 
\rd^{3|4}z\, E \,
\bigg\{\frac{1}{8}U^{\a(2s-2)}\cD^{\g}({\bar \cD}^{2}-6\mu)\cD_{\g}U_{\a(2s-2)}\non \\
&&+2s(s+1)\mu\bar{\mu} U^{\a(2s-2)}U_{\a(2s-2)} \non \\
&&+\frac{s}{2s+1}U^{\a(2s-2)}\Big(\cD^{\b} {\bar \cD}^{\g}
V_{\b \g \a(2s-2) }
-{\bar \cD}^{\b}{\cD}^{\g}\bar{V}_{\b \g \a(2s-2)} \Big) \non \\ 
&& +\frac{s}{2s-1}  \bar{V}^{\a(2s)} V_{\a(2s)} 
+ \frac{s}{2(2s+1)}\Big(V^{\a(2s)}V_{\a(2s)}+\bar{V}^{\a(2s)} \bar V_{\a(2s)}\Big)
 \non \\
&& + \frac{4s}{(2s+1)(2s-1)} \Big(V^{\a(2s)} \G_{\a(2s)}
+ \bar V^{\a(2s)} \bar \G_{\a(2s)}\Big) \bigg\}~,
\label{510}
\eea
in which $V_{\a(2s)} $ is an unconstrained complex superfield, 
and $\G_{\a(2s)}$ is given by \eqref{5.8}. This action is 
invariant under the gauge transformation  \eqref{H-gauge-int}
accompanied with
\begin{subequations}
\bea
\delta_L V_{\a(2s)}&=&\delta_L G_{\a(2s)}~,\\
\d_L \G_{\a(2s)}&=&
-\frac{1}{4}(\bar{\cD}^{2} +4s\mu)\cD_{(\a_{1}}\bar{L}_{\a_2 \dots \a_{2s})}
+  \frac{\ri}{2}  (2s+1)\bar{\cD}^{\g}\cD_{(\g\a_{1}}\bar{L}_{\a_2 \dots \a_{2s})} \non \\
&=&
\frac{1}{2}\cD_{(\alpha_{1}}\bar{\cD}_{\alpha_{2}}{\gamma}_{\a_3 \dots \a_{2s})}-\frac{\ri}{2} (2s-1)
 \cD_{(\alpha_{1}\alpha_{2}}{\gamma}_{\a_3 \dots \a_{2s})} 
~,
\label{tr-gauge-int}
\eea
\end{subequations}
where ${\gamma}_{\a(2s-2)} =-\bar{\cD}^{\b}\bar{L}_{\b \a_1 \dots \a_{2s-2}} $,
and $\delta_L G_{\a(2s)}$ is given by \eqref{int-long-gauge}.
The first-order model described by action \eqref{510} is equivalent to 
the longitudinal theory \eqref{long-action-int}. Indeed, varying 
$S_{s} [U,V, \bar V, \G, \bar \G] $ with respect to 
the prepotential $\Phi_{\a(2s+1)}$, eq. \eqref{5.8}, gives 
$ V_{\a(2s)}= G_{\a(2s)}$, and then the action  \eqref{510} 
reduces to  the longitudinal one, eq. \eqref{long-action-int}.
On the other hand, we can integrate out 
the auxiliary superfield $V_{\a(2s)}$ and its conjugate $\bar V_{\a(2s)}$
from \eqref{tr-gauge-int} 
using their equations of motion. This leads to the transverse action
\bea
S_{(s)}^{\perp}
&=& \Big(-\hf\Big)^{s}\int 
\rd^{3|4}z
\, E \,
\bigg\{\frac{1}{8}U^{\a(2s-2)}\cD^{\g}({\bar \cD}^{2}-6\mu)\cD_{\g}U_{\a(2s-2)}\non \\
&& -\frac{2s-1}{16(2s+1)} \Big( 
8s  {\cD}^{\a_1 \a_2} U^{\a_3 \dots \a_{2s}} {\cD}_{(\a_1 \a_2} U_{\a_3 \dots \a_{2s})} \non \\
&& + [\cD^{\a_1}, {\bar \cD}^{\a_2}]U^{\a_3 \dots \a_{2s}} 
[\cD_{(\a_1}, {\bar \cD}_{\a_2}]U_{\a_3 \dots \a_{2s})}\Big) \non \\
&& 
+2s(s+1)\mu\bar{\mu} U^{\a(2s-2)}U_{\a(2s-2)}
-\ri U^{\a_1 \dots \a_{2s-2}} {\cD}^{\a_{2s-1}  \a_{2s}} 
\big( \G_{\a(2s)} -\bar \G_{\a(2s)} \big)\non \\
&&- \frac{2}{2s-1} \bar \G^{\a(2s)} \G_{\a(2s)} 
+\frac{1}{2s+1} (\G^{\a(2s)} \G_{\a(2s)}+\bar{\G}^{\a(2s)} \bar \G_{\a(2s)})
 \bigg\}~.
\label{tr-action-int}
\eea
The action  is invariant under \eqref{H-gauge-int} and \eqref{tr-gauge-int}.


\subsection{Reformulation of the longitudinal theory }\label{subsection5.3} 

In this subsection we consider a reformulation of the longitudinal theory 
that is similar to the one proposed in the 4D $\cN=1$ AdS case \cite{BHK}.
It is obtained by enlarging the gauge freedom \eqref{5.4ab}
at the cost of introducing new purely gauge superfield variables in addition to 
$U_{\a(2s-2)}$, $\J_{\a(2s-1)}$ and $\bar\J_{\a(2s-1)}$.
In such a setting, the  gauge freedom of  $\J_{\a(2s-1)}$ 
coincides with that of a {\it complex} conformal  gauge superfield.
Given a positive integer $s \geq 2$, a massless superspin-$s$ multiplet can be described in ${\rm AdS}^{(3|1,1)}$
by using the following superfield variables: 
(i) an unconstrained prepotential
 $\J_{\a(2s-1)}  $ 
and its complex conjugate $\bar \J_{\a(2s-1)}$; 
(ii) a real  superfield 
 $U_{\a(2s-2)} =\bar U_{\a(2s-2)}  $; and (iii)
a complex superfield $\S_{\a(2s-3)}$ 
and its conjugate $\bar \S_{\a (2s-3)}$, 
where $\S_{\a(2s-3) }$ is constrained 
to be transverse linear,
\bea
\bar \cD^\b \S_{\b \a(2s-4)} =0~.
\label{2.1}
\eea
The constraint \eqref{2.1} is solved in terms of an unconstrained complex  prepotential $Z_{\a(2s-2)}$ by the rule
\bea
\S_{\a(2s-3)} = \bar \cD^\b Z_{ (\b \a_1 \dots \a_{2s-3})} ~.
\label{2.2}
\eea
This prepotential is defined modulo gauge transformations 
\bea
\d_\x Z_{\a(2s-2)}=  \bar \cD^\b \x_{(\b \a_1 \dots \a_{2s-2})} ~,
\label{2.3}
\eea
with the gauge parameter  $\x_{\a(2s-1)}$ being  unconstrained. 
 
The gauge freedom of $\J_{\a_1 \dots \a_{2s-1}} $ is given by
\begin{subequations} \label{2.4}
\bea
 \d_{ {\mathfrak V} ,\z} \J_{\a_1 \dots \a_{2s-1}} 
 &=&  \cD_{(\a_1}  {\mathfrak V}_{\a_2 \dots \a_{2s-1})}
+  \bar \cD_{(\a_1} \z_{\a_2 \dots \a_{2s-1} )}  ~ , \label{2.4a}
\eea
with unconstrained gauge parameters ${\mathfrak V}_{\a(2s-2)}$ 
and $\z_{\a(2s-2)}$. 
The $\mathfrak V$-transformation is defined to act on the superfields 
$U_{\a(2s-2)}$ and $\S_{\a(2s-3)}$ as follows
\bea
\d_{\mathfrak V} U_{\a(2s-2)}&=& {\mathfrak V}_{\a(2s-2)} 
+\bar {\mathfrak V}_{\a(2s-2)}
~, \label{2.4b}\\
\d_{\mathfrak V} \S_{\a(2s-3) }&=&  \bar \cD^\b \bar {\mathfrak V}_{\b \a(2s-3)}
\quad \Longrightarrow \quad \d_{\mathfrak V} Z_{\a(2s-2)}
=\bar  {\mathfrak V}_{\a(2s-2) }~.~~~
\label{2.4c}
\eea
\end{subequations}
The longitudinal linear superfield defined by \eqref{g2.5}
is invariant under the $\z$-transformation \eqref{2.4a} 
and  varies under the $\mathfrak V$-transformation as 
\bea
 \d_{ {\mathfrak V} } G_{\a_1 \dots \a_{2s}} 
 &=&  \bar \cD_{(\a_1} \cD_{\a_2}  {\mathfrak V}_{\a_3 \dots \a_{2s})}~.
\eea

The gauge-invariant action is given by 
\bea
S^{\|}_{(s)} &=&
\Big( - \frac{1}{2}\Big)^s  \int 
\rd^{3|4}z\, E
\left\{ \frac{1}{8} U^{ \a (2s-2) }  \cD^\b ({\bar \cD}^2- 6\mu) \cD_\b 
U_{\a (2s-2)} \right. \non \\
&&+ \frac{s}{2s+1}U^{ \a(2s-2) }
\Big( \cD^{\b}  {\bar \cD}^{\g} G_{\b \g \a(2s-2)}
- {\bar \cD}^{\b}  \cD^{\g} 
{\bar G}_{\b \g \a(2s-2) } \Big) \non \\
&&+ 2s(s+1) \bar \mu \mu U^{\a (2s-2)} U_{\a (2s-2)} \non \\
&&+ \frac{s}{2s-1} \bar G^{ \a (2s)} G_{ \a (2s)} 
+ \frac{s}{2(2s+1)}\Big( G^{ \a (2s) } G_{ \a (2s)} 
+ \bar G^{ \a (2s) }  \bar G_{ \a (2s) } 
 \Big) \non \\
 &&+ \hf \frac{s-1}{2s-1}U^{ \a(2s-2) }
\Big( \cD_{\a_1} \bar \cD^2 \bar \S_{\a_2 \dots \a_{2s-2}}
 - {\bar \cD}_{\a_1}  \cD^2 \S_{\a_2 \dots \a_{2s-2} } \Big)  \non \\
&&+\frac{1}{2s-1} \J^{\a(2s-1)} \Big( 
\cD_{\a_1} \bar \cD_{\a_2} -2\ri (s-1) \cD_{\a_1 \a_2} \Big)
\S_{\a_3 \dots \a_{2s-1} }\non  \\
&&+\frac{1}{2s-1} \bar \J^{\a(2s-1)} \Big( 
\bar \cD_{\a_1} \cD_{\a_2} -2\ri (s-1) \cD_{\a_1 \a_2} \Big)
\bar \S_{\a_3 \dots \a_{2s-1} }\non  \\
&&- \mu (s+3) U^{\a (2s-2)} \cD_{\a_1} 
\bar \S_{\a_2 \dots \a_{2s-2}} + \bar \mu (s+3) U^{\a (2s-2)} \bar \cD_{\a_1} 
 \S_{ \a_2 \dots \a_{2s-2}} \non \\
&&+ \frac{s-1}{4(2s-1)} \Big( \S^{\a(2s-3) } \cD^2 \S_{\a(2s-3)} 
- \bar \S^{\a(2s-3) }\bar \cD^2 \bar \S_{\a(2s-3)} \Big) \non \\
&&- \frac{1}{2s-1}\bar \S^{\a(2s-3)} \Big( (2s^2-s+1) \cD^\b \bar \cD_{\a_1 }
+2\ri \frac{ (s-1)(2s-3)}{2s-1} \cD^\b{}_{\a_1} \Big) \S_{\b \a_2 \dots \a_{2s-3}} \non\\
&&+ \mu (s+3) \bar \S^{\a(2s-3)} \bar \S_{\a(2s-3)} + \bar \mu (s+3) \S_{\a(2s-3)}  \S^{\a(2s-3)} \Big\}
~,~~~
\label{action}
\eea
with gauge symmetries \eqref{2.4} and, by construction, \eqref{2.3}. The above action is real due to the identity \eqref{1.44}.

The $\mathfrak V$-gauge freedom \eqref{2.4} allows us to gauge away $\S_{\a(2s-3)}$,
\bea
 \S_{\a(2s-3)}=0~.
\label{2.10}
\eea
In this gauge, the action \eqref{action} reduces to 
that describing the longitudinal formulation for the massless superspin-$s$ multiplet  \eqref{long-action-int}. The gauge condition \eqref{2.10} does not 
fix completely the $\mathfrak V$-gauge freedom. The residual gauge transformations  
are generated by 
\bea
{\mathfrak V}_{\a(2s-2)} = \cD^\b L_{(\b \a_1 \dots \a_{2s-2})}~,
\label{2.11}
\eea
with $L_{\a(2s-2)}$ being an unconstrained superfield. With this expression for ${\mathfrak V}_{\a(2s-2)}$, the gauge transformations \eqref{2.4a}  and \eqref{2.4b} coincide with \eqref{int-long-gauge}.
Thus, the action \eqref{action} indeed provides an off-shell formulation for the massless superspin-$s$ multiplet in (1,1) AdS superspace.

The action \eqref{action} includes a single term which involves the `naked' 
gauge field $\bar \J_{\a(2s-1)} $ and not the field strength $\bar G_{\a(2s)} $, 
the latter being 
defined by \eqref{g2.5} and invariant under the $\z$-transformation \eqref{2.4a}.
This is actually a $BF$ term, for it can be written in two different forms
\bea
 \int  
 \rd^{3|4}z \, E
 \,
 \bar \J^{\a(2s-1)} \Big( 
 \bar \cD_{\a_1} \cD_{\a_2} &-&2\ri (s-1) \cD_{\a_1 \a_2} \Big)
\S_{\a_3 \dots \a_{2s-1} } \non \\
= -\frac{2s}{2s+1}  
\int 
\rd^{3|4}z \, E
\,
 \bar G^{\a(2s)} \Big( \cD_{\a_1} \bar \cD_{\a_2}  
&+&\ri (2s+1) \cD_{\a_1 \a_2} \Big)
\bar Z_{\a_3 \dots \a_{2s} }~.~~~
\label{2.14}
\eea
The former makes the gauge symmetry \eqref{2.3} manifestly realised, 
while the latter
turns the $\z$-transformation \eqref{2.4a} into a manifest symmetry.

Making use of \eqref{2.14} leads to a different representation 
for the action \eqref{action}. It is 
\bea
S^{\|}_{(s)} &=&
\Big( - \frac{1}{2}\Big)^s  \int 
\rd^{3|4}z
\, E
\left\{ \frac{1}{8} U^{ \a (2s-2) }  \cD^\b ({\bar \cD}^2- 6\mu) \cD_\b 
U_{\a (2s-2)} \right. \non \\
&&+ \frac{s}{2s+1}U^{ \a(2s-2) }
\Big( \cD^{\b}  {\bar \cD}^{\g} G_{\b \g \a(2s-2)}
- {\bar \cD}^{\b}  \cD^{\g} 
{\bar G}_{\b \g \a(2s-2) } \Big) \non \\
&&+ 2s(s+1) \bar \mu \mu U^{\a (2s-2)} U_{\a (2s-2)} \non \\
&&+ \frac{s}{2s-1} \bar G^{ \a (2s)} G_{ \a (2s)} 
+ \frac{s}{2(2s+1)}\Big( G^{ \a (2s) } G_{ \a (2s)} 
+ \bar G^{ \a (2s) }  \bar G_{ \a (2s) } 
 \Big) \non \\
 &&+ \hf \frac{s-1}{2s-1}U^{ \a(2s-2) }
\Big( \cD_{\a_1} \bar \cD^2 \bar \S_{\a_2 \dots \a_{2s-2}}
 - {\bar \cD}_{\a_1}  \cD^2 \S_{\a_2 \dots \a_{2s-2} } \Big)  \non \\
&&+\frac{2s}{(2s-1)(2s+1)} G^{\a(2s)} \Big( \bar \cD_{\a_1} \cD_{\a_2} +\ri (2s+1) \cD_{\a_1 \a_2} \Big)
Z_{\a_3 \dots \a_{2s} }\non  \\
&&-\frac{2s}{(2s-1)(2s+1)} \bar G^{\a(2s)} \Big( \cD_{\a_1} \bar \cD_{\a_2} +\ri (2s+1) \cD_{\a_1 \a_2} \Big)
\bar Z_{\a_3 \dots \a_{2s} }\non  \\
&&- \mu (s+3) U^{\a (2s-2)} \cD_{\a_1} 
\bar \S_{\a_2 \dots \a_{2s-2}} + \bar \mu (s+3) U^{\a (2s-2)} \bar \cD_{\a_1} 
 \S_{ \a_2 \dots \a_{2s-2}} \non \\
&&+ \frac{s-1}{4(2s-1)} \Big( \S^{\a(2s-3) } \cD^2 \S_{\a(2s-3)} 
- \bar \S^{\a(2s-3) }\bar \cD^2 \bar \S_{\a(2s-3)} \Big) \non \\
&&- \frac{1}{2s-1}\bar \S^{\a(2s-3)} \Big( (2s^2-s+1) \cD^\b \bar \cD_{\a_1 }
+2\ri \frac{ (s-1)(2s-3)}{2s-1} \cD^\b{}_{\a_1} \Big) \S_{\b \a_2 \dots \a_{2s-3}} \non\\
&&+ \mu (s+3) \bar \S^{\a(2s-3)} \bar \S_{\a(2s-3)} + \bar \mu (s+3) \S_{\a(2s-3)}  \S^{\a(2s-3)} \Big\}
~.~~~
\label{action2}
\eea

Before concluding this section, it is worth discussing the 
structure of the dynamical variable $\J_{\a(2s-1)}$. 
This superfield is unconstrained complex, and its gauge transformation law
is given by eq. \eqref{2.4a}. Comparing \eqref{2.4a} 
with the gauge transformation law \eqref{2.28}  $n =2s-1$, 
which corresponds to  the conformal gauge superfield ${\mathfrak H}_{\a(2s-1)} $, 
we see that $\J_{\a(2s-1)}$ may be interpreted as a complex 
conformal gauge superfield.


\subsection{Massless gravitino multiplet}\label{subsection5.4}

The massless gravitino multiplet, which corresponds to the $s=1$ case, 
was excluded from our consideration of the previous subsection. 
Here we will fill the gap.

The (generalised) longitudinal formulation for the gravitino 
multiplet is described by the action 
\bea
S^{\|}_{\rm GM} &=& -\hf \int \rd^{3|4}z
\, E\,
\bigg\{ \frac{1}{8} U  \cD^\b ({\bar \cD}^2- 6\mu) U
+ \frac{1}{3}U\big( \cD^{\a}  {\bar \cD}^{\b} G_{\a \b}
- {\bar \cD}^{\a}  \cD^{\b} {\bar G}_{\a \b } \big)  \non \\
&& + \bar G^{\a\b} G_{\a\b} 
+\frac 16 \big( G^{\a\b} G_{\a\b} +\bar G^{\a\b} \bar G_{\a\b} \big) \non \\
&&+|\m|^2 \Big(2U - \frac{\F}{\m} -\frac{\bar \F}{\bar \m} \Big)^2
+2\Big( \frac{\F}{\m} +\frac{\bar \F}{\bar \m}\Big)
\Big( \m \cD^\a \J_\a + \bar \m \bar \cD_\a \bar \J^\a \Big) 
\bigg\}~,
\label{5.21}
\eea
where $\F$ is a covariantly chiral scalar superfield, $\bar \cD_\a \F=0$, and
\bea
G_{\a\b} = \bar \cD_{(\a} \J_{\b)} ~, \qquad 
\bar G_{\a\b} = -\cD_{(\a} \bar \J_{\b)} ~.
\eea
  This action is invariant under gauge transformations of the form 
\begin{subequations}\label{5.23}
\bea
\d U&=& {\mathfrak V} +\bar {\mathfrak V}~,  \label{3.28a} \\
\d \J_\a &=& =  \cD_\a  {\mathfrak V}+ \bar \cD_\a \z~, 
 \\
\d \F &=& -\frac 14 (\bar \cD^2 -4\m) \bar  {\mathfrak V}~,
\label{5.23c} 
\eea
\end{subequations}
where the gauge parameters $\mathfrak V$ and $\z$ are unconstrained complex
superfields. 
  
The gauge $\mathfrak V$-freedom \eqref{5.23} allows us to
 impose the condition $\F=0$. In this gauge 
the action \eqref{5.21}  turns into  \eqref{long-action-int} with $s=1$, and the residual 
gauge $\mathfrak V$-freedom is described by 
${\mathfrak V} = \cD^\b L_\b$, where the spinor
gauge parameter $L_\a$ is unconstrained complex.
  
The action \eqref{5.21} involves the chiral scalar $\F$ and its conjugate only in the combination $(\vf + \bar \vf)$, where $\vf = \F /\m$. This means that the model 
\eqref{5.21} possesses a dual formulation realised in terms of a real linear superfield
subject to the constraint
\eqref{2.22}.
  

\section{Higher-spin supercurrents}

Inspired by the analysis of Dumitrescu and Seiberg \cite{DS},
the most general supercurrent multiplets for theories with 
(1,1) AdS or (2,0) AdS supersymmetry were introduced in \cite{KT-M11},
with the (1,1) AdS case being 
a natural extension of the 4D $\cN=1 $ AdS supercurrents classified in 
\cite{BK12,BK11}. Here we will formulate higher-spin supercurrents 
in (1,1) AdS superspace by making use of the off-shell formulations
for massless supersymmetric higher-spin gauge theories in (1,1) AdS superspace, 
which have been constructed in the previous two sections. 
Our analysis will be analogous to the one recently  given in the 4D $\cN=1$ case \cite{BHK}. 


\subsection{Non-conformal supercurrents: Half-integer superspin} 

The two off-shell formulations for the massless supers[in-$(s+\hf)$ multiplet, 
which we reviewed in sections  \ref{subsection4.1} and  \ref{subsection4.2}, 
lead to different higher-spin supercurrent multiplets. 
In this subsection we first described the explicit structure of these supermultiplets 
and then show how they are related to each other. 


\subsubsection{Longitudinal supercurrent}

In the framework of the longitudinal formulation \eqref{long-action-half}, let us couple the prepotentials 
${\mathfrak H}_{ \a (2s)} $, $\J_{ \a (2s-3)}$ and $\bar \J_{ \a (2s-3)}  $,
to external sources
\bea
S^{(s+\hf)}_{\rm source}= \int 
\rd^{3|4}z\, E\, \Big\{ 
{\mathfrak H}^{ \a (2s)} J_{ \a (2s)}
+ \J^{ \a (2s-3)} T_{ \a (2s-3)}
+ \bar \J_{ \a (2s-3) } \bar T^{ \a (2s-3)} \Big\}~.
\label{7.1}
\eea
Requiring $S^{(s+\hf)}_{\rm source}$ to be invariant under 
\eqref{long-prep-gauge} gives
\begin{subequations} \label{7.3}
\bea
\bar \cD^\b T_{\b \a(2s-4)} =0~,
\label{7.2}
\eea
and therefore $T_{ \a (2s-3)} $ is a transverse linear superfield. 
Requiring $S^{(s+\hf)}_{\rm source}$ to be invariant under the gauge transformations
(\ref{H-gauge}) and (\ref{prep-gauge-L}) gives the following conservation equation:
\bea
\bar \cD^\b J_{\b \a(2s-1)} 
+ \hf \Big( \cD_{(\a_1} \bar \cD_{\a_2}
-2\ri (s-1) \cD_{ (\a_1 \a_2 } \Big)  T_{\a_3\dots \a_{2s-1})} =0~.
\label{7.3a}
\eea
For completeness, we also give the conjugate equation
\bea
\cD^\b J_{\b \a(2s-1)} 
- \hf \Big( \bar \cD_{(\a_1}  \cD_{\a_2}
-2\ri (s-1) \cD_{ (\a_1 \a_2 } \Big)  \bar T_{\a_3 \dots \a_{2s-1})} 
=0~. \label{7.3c}
\eea
\end{subequations}

As in \cite{BHK}, it is useful to introduce 
auxiliary real variables $\z^\a$. Given a tensor superfield $U_{\a(m)}$, we associate with it 
the following  field
\bea
U_{(m)} (\z):= \z^{\a_1} \dots \z^{\a_m} U_{\a_1 \dots \a_m}~,
\label{4.100}
\eea
which is homogeneous of degree $m$ in the variables $\z^\a$.
We introduce operators that  increase the degree 
of homogeneity in the variable $\z^\a$, 
\begin{subequations}
\bea
{\cD}_{(1)} &:=& \z^\a \cD_\a~,\\
{\bar \cD}_{(1)} &:=&  \z^\a \bar \cD_\a~, \\
{\cD}_{(2)} &:=& \ri \z^\a \z^\b \cD_{\a\b}
= -\hf\big\{ {\cD}_{(1)} , \bar {\cD}_{(1)} \big\}
~.
\eea
\end{subequations}
We also introduce two operators that decrease the degree 
of homogeneity in the variable $\z^\a$, specifically
\begin{subequations}
\bea
\cD_{(-1)} &:=& \cD^\a \frac{\pa}{\pa \z^\a}~,\\
\bar \cD_{(-1)}& :=& \bar \cD^\a \frac{\pa}{\pa  \z^\a}~ 
~.
\eea
\end{subequations}

Making use of the above notation, the transverse linear condition \eqref{7.2} and its conjugate become
\begin{subequations}
\bea
\bar \cD_{(-1)} T_{(2s-3)} &=&0~,  \label{7.8a}\\
\cD_{(-1)} \bar T_{(2s-3)} &=&0~.  \label{7.8b}
\eea
\end{subequations}
The conservation equations \eqref{7.3a} and \eqref{7.3c} turn into 
\begin{subequations}
\bea
\frac{1}{2s}\bar \cD_{(-1)} J_{(2s)} -\hf A_{(2)} T_{(2s-3)}&=&0~, \label{7.9a}\\
\frac{1}{2s}\cD_{(-1)} J_{(2s)} -\hf \bar A_{(2)} \bar T_{(2s-3)}&=&0~. \label{7.9b}
\eea
\end{subequations}
where 
\bea
A_{(2)} := -\cD_{(1)} \bar \cD_{(1)} + 2(s-1) \cD_{(2)} ~, \quad
\bar A_{(2)} := \bar \cD_{(1)}  \cD_{(1)} -2(s-1) \cD_{(2)} ~. 
\eea
Since 
$(\bar \cD_{(-1)}) ^2 J_{(2s)} =0$,
the conservation equation \eqref{7.9a} is consistent provided
\bea
\bar \cD_{(-1)}  A_{(2)} T_{(2s-3)}=0~.
\eea
This is indeed true, as a consequence of the transverse linear condition
\eqref{7.8a}. 

\subsubsection{Transverse supercurrent}
One can also make use of the transverse formulation \eqref{tr-action-half} and couple the prepotentials ${\mathfrak H}_{ \a (2s)} $, $\Phi_{ \a (2s-1)}$ and $\bar \Phi_{ \a (2s-1)}  $ to external sources
\bea
S^{(s+\hf), tr}_{\rm source}= \int 
\rd^{3|4}z\, E\, \Big\{ 
{\mathfrak H}^{ \a (2s)} {\mathbb J}_{ \a (2s)}
+ \Phi_{ \a (2s-1)} \bar {\mathbb F}^{ \a (2s-1)}
+ \bar \Phi^{ \a (2s-1) } {\mathbb F}_{ \a (2s-1)} \Big\}~.
\label{source-tr}
\eea
Requiring that the action \eqref{source-tr} be invariant under the gauge transformations \eqref{H-gauge}, \eqref{eee1}, and \eqref{tr-prep-gauge} leads to the following conditions on the transverse supercurrent multiplet 
\begin{subequations} \label{tr-current}
\bea
\bar \cD_{(\a_1} \bar {\mathbb F}_{\a_2 \dots \a_{2s})} &=&0~, \label{eee2} \\
\bar \cD^{\b} {\mathbb J}_{\b \a(2s-1)}-\frac {1}{4} (\bar \cD^2 + 2\m (2s-1)) {\mathbb F}_{\a(2s-1)} &=&0~. \label{eee3}
\eea
\end{subequations}
Thus, the trace multiplet $\bar {\mathbb F}_{\a(2s-1)}$ is longitudinal linear. 

\subsubsection{Improvement transformation}
We now construct a well-defined improvement transformation which converts the higher-spin supercurrent \eqref{7.3} to \eqref{tr-current}, thus showing that they are indeed equivalent. 

The transverse linearity condition \eqref{7.2} implies that there exists a well-defined complex tensor operator  $X_{\a(2s-2)}$ such that 
\bea
T_{\a(2s-3)}
= \bar \cD^\b X_{\b \a(2s-3)} ~.
\eea
Let us split $X_{\a(2s-2)}$ into its real and imaginary parts, 
\bea
X_{\a(2s-2)} = U_{\a(2s-2)} + \ri V_{\a(2s-2)}~.
\eea
Then one may check that the operators
\begin{subequations} \label{614}
\bea
{\mathbb J}_{\a(2s)} &:=& J_{\a(2s)}
+\frac{s}{2} \big[ \cD_{(\a_1}, \bar \cD_{\a_2} \big]
U_{\a_2 \dots \a_{2s-1} ) }
+ s \cD_{(\a_1 \a_2 } V_{\a_3 \dots \a_{2s})}~, ~~~\\
{\mathbb F}_{\a(2s-1)} &:=& 
\cD_{(\a_1} \Big\{ 2s U_{\a_2 \dots \a_{2s-1)}}
- \ri V_{\a_2 \dots \a_{2s-1})}\Big\} \eea
\end{subequations}
satisfy the conservation equation \eqref{eee3} and the longitudinal linear condition \eqref{eee2}.

The improvement transformation \eqref{614}
turns the higher-spin supercurrent \eqref{7.3} to \eqref{tr-current}
It is also not difficult to construct an inverse improvement transformation 
converting  the higher-spin supercurrent \eqref{tr-current} to \eqref{7.3}.
Therefore the higher-spin supercurrents \eqref{7.3} and \eqref{tr-current}
are equivalent, and it is suffices to work with one of them, say, 
the longitudinal supermultiplet \eqref{7.3}.
The situation proves to be analogous in the integer superspin case, 
for which we will formulate in the next subsection a higher-spin supercurrent
associated with  the new gauge formulation \eqref{action}.


\subsection{Non-conformal supercurrents: Integer superspin} \label{subsection4.4}

We now make use of the new gauge formulation \eqref{action}, 
or  equivalently \eqref{action2}, for the integer superspin-$s$ multiplet to derive the 3D analogue of the non-conformal higher-spin supercurrents proposed in \cite{BHK}.

Let us couple the prepotentials 
$U_{ \a (2s-2) } $, $Z_{ \a (2s-2) }$ and $\Psi_{ \a (2s-1) } $ to external sources
\bea
S^{(s)}_{\rm source} &=& \int 
\rd^{3|4}z
\, E\, \Big\{ 
\Psi^{ \a (2s-1)  } J_{ \a (2s-1) }
-\bar \Psi^{ \a (2s-1) } \bar J_{ \a (2s-1) }
+U^{ \a (2s-2)  } S_{ \a (2s-2)  } \non \\
&&\qquad \qquad ~~
+ Z^{ \a (2s-2) } T_{ \a (2s-2) } 
+ \bar Z^{ \a (2s-2)  } \bar T_{ \a (2s-2) }
 \Big\}~.
\label{4.1}
\eea
In order for $S^{(s)}_{\rm source}$ to be invariant under the $\z$-transformation 
in \eqref{2.4a}, the source  $J_{ \a (2s-1) }$ must satisfy
\bea
\bar \cD^\b J_{\b \a(2s-2)} =0 \quad \Longleftrightarrow \quad
\cD^\b \bar J_{\b \a(2s-2)} =0 ~.
\label{4.2a}
\eea
Next, requiring $S^{(s)}_{\rm source}$ to be invariant under the transformation \eqref{2.3} leads to
\bea
\bar \cD_{(\a_1} T_{\a_2 \dots \a_{2s-1})} =0
 \quad \Longleftrightarrow \quad
 \cD_{(\a_1} \bar T_{\a_2 \dots \a_{2s-1})} =0~.
\label{4.2b}
\eea
We see that  the superfields $J_{ \a (2s-1) }$ and $T_{ \a (2s-2) } $ are transverse linear and longitudinal linear, respectively.
Finally, requiring $S^{(s)}_{\rm source}$ to be invariant under the 
$\mathfrak V$-transformation 
\eqref{2.4} gives the following conservation equation
\begin{subequations} \label{3.4}
\bea
- \cD^\b J_{\b \a(2s-2)} 
+S_{\a(2s-2)} + \bar T_{\a(2s-2) } =0
\label{4.2c}
\eea
as well as its conjugate
\bea
 \bar \cD^\b \bar J_{ \b \a(2s-2)} 
+S_{\a(2s-2)} + T_{\a(2s-2)} =0~. 
\label{3.4b}
\eea
\end{subequations}

Taking the sum of \eqref{4.2c} and \eqref{3.4b}
leads to
\bea
 \cD^\b J_{\b \a(2s-2)} 
+ \bar \cD^\b \bar J_{\b \a(2s-2)}
+ T_{\a(2s-2)}-\bar T_{\a(2s-2)} =0~. 
\label{4.3}
\eea
As a consequence of \eqref{4.2b}, the conservation equation \eqref{4.3} 
implies
\bea
 \cD_{(\a_1} \left\{\cD^{|\b|} J_{\a_2 \dots \a_{2s-1}) \b} 
+ \bar \cD^\b \bar J_{ \a_2 \dots \a_{2s-1}) \b}\right\}
+\cD_{(\a_1} T_{\a_2 \dots \a_{2s-1})} =0~. 
\label{4.4}
\eea

Using our notation introduced in the previous subsection, 
the transverse linear condition \eqref{4.2a} turns into 
\bea
\bar \cD_{(-1)} J_{(2s-1)} &=& 0~,  \label{4.5a}
\eea
while the longitudinal linear condition \eqref{4.2b} takes the form
\bea
\bar \cD_{(1)} T_{(2s-2)} &=& 0~. \label{4.5b}
\eea
The conservation equation \eqref{4.2c} becomes
\bea
-\frac{1}{(2s-1)} \cD_{(-1)} J_{(2s-1)} + S_{(2s-2)} + \bar T_{(2s-2)} = 0
\label{4.6}
\eea
and \eqref{4.4} takes the form
\bea
\frac{1}{(2s-1)} \cD_{(1)} \left\{\cD_{(-1)} J_{(2s-1)} + \bar \cD_{(-1)} \bar J_{(2s-1)}\right\}
+\cD_{(1)} T_{(2s-2)} =0~. 
\label{4.7}
\eea


\section{Higher-spin supercurrents for chiral matter: Half-integer superspin}

In the remainder of this paper we will study explicit realisations
of the higher-spin supercurrents introduced above in supersymmetric 
field theories in AdS. 

\subsection{Superconformal model for a chiral superfield} \label{1chiral}


Let us consider the superconformal theory of a single chiral scalar superfield 
\bea
S = \int 
\rd^{3|4}z
\,E\, \bar \F \F ~,
\label{chiral}
\eea
where  $\F$ is covariantly chiral, $\bar \cD_\a \F =0$.
We construct the following conformal supercurrent $J_{(2s)}$, which is a minimal extension of the conserved supercurrent constructed in flat ${\cal N}=2$ Minkowski superspace \cite{NSU}.
\bea
J_{(2s)} &=& \sum_{k=0}^s (-1)^k
\left\{ \hf \binom{2s}{2k+1} 
{\cD}^k_{(2)} \bar \cD_{(1)} \bar \F \,\,
{\cD}^{s-k-1}_{(2)} \cD_{(1)} \F  
+ \binom{2s}{2k} 
{\cD}^k_{(2)} \bar \F \,\, {\cD}^{s-k}_{(2)} \F \right\}~.~~~
\label{7.15}
\eea
Making use of
the massless equations of motion,   $(\cD^2-4\bar \m)\, \F = 0$, 
one may check that $J_{(2s)}$ satisfies the conservation equation
\bea
\cD_{(-1)} J_{(2s)} = 0 \quad \Longleftrightarrow \quad 
\bar \cD_{(-1)} J_{(2s)} = 0 ~.~
\label{7.9}
\eea
The calculation of \eqref{7.9} in AdS is much more complicated than in flat superspace due to the fact that the algebra of covariant derivatives \eqref{1.2}
is nontrivial. 
Let us sketch the main steps in evaluating the left-hand side of eq.~\eqref{7.9}
with $J_{(2s)} $ given by \eqref{7.15}.
We start with the obvious relations
\begin{subequations}
\bea
\frac{\pa}{\pa \z^\a} {\cD}_{(2)} &=&2\ri { \z}^\b {\cD}_{\a \b}~, \\
\frac{\pa}{\pa \z^\a} {\cD}^k_{(2)} &=& 
\sum_{n=1}^k\,{\cD}^{n-1}_{(2)} \,\,  2\ri \, {\z}^\b {\cD}_{\a \b}\,\, {\cD}^{k-n}_{(2)} ~, \qquad k>1
~.\label{eq1}
\eea
\end{subequations}
To simplify eq.~\eqref{eq1}, we may push ${\z}^\b{\cD}_{\a \b}$, say,  to the left 
provided that we take into account its commutator with ${\cD}_{(2)}$:
\bea
[{ \z}^\b {\cD}_{\a \b}\,, {\cD}_{(2)}] = -4\ri \,\bar \m \m \,\z_\a  { \z}^\b{\z}^\g {M}_{\b \g}~.
\label{555}
\eea
Associated with the Lorentz generators are the operators
\bea
{M}_{(2)} &:=& {\z}^\a {\z}^\b {M}_{\a \b}~,
\eea
where ${M}_{(2)}$ appears in the right-hand side of \eqref{555}.
These operators annihilate every superfield $U_{(m)}(\z) $ of the form 
\eqref{4.100},
\bea
{M}_{(2)} U_{(m)} =0~.
\eea
From the above consideration, it follows that
\begin{subequations}
\bea
[{\z}^\b {\cD}_{\a \b}\,, {\cD}^k_{(2)}]\, U_{(m)} &=& 0 ~, \\
\Big(\frac{\pa}{\pa \z^\a} {\cD}^k_{(2)}\Big)U_{(m)} &=& 2\ri k\, {\z}^\b {\cD}_{\a \b}\, {\cD}^{k-1}_{(2)}U_{(m)}~.
\eea
\end{subequations}
We also state some other properties which we often use throughout our calculations
\begin{subequations}
\bea
{\cD}^2_{(1)} &=& -2\bar \m M_{(2)} ~,\\
\big[ {\cD}_{(1)}\,, {\cD}_{(2)} \big] 
&=& 
\big[ \bar \cD_{(1)}\,, \cD_{(2)} \big] = 0~,\\
\big[ \cD^\a, \cD_{(2)} \big] &=& -2 \bar \m \, \z^\a \bar {\cD}_{(1)} ~,\\
\big[\cD^\a, \cD^k_{(2)}\big] &=& -2 \bar \m \,k \,\z^\a \cD^{k-1}_{(2)} \bar {\cD}_{(1)}~,\\
\big[\cD^\a, \z^\b \cD_{\a \b}\big] &=& 3 \ri \bar \m \,\bar \cD_{(1)}~.
\eea
\end{subequations}
The above identities suffice to prove that the supercurrent  \eqref{7.15}
does obey the conservation equation \eqref{7.9}.


\subsection{Non-superconformal model for a chiral superfield} \label{subsection5.2}

Let us now add the mass term to~\eqref{chiral} and consider the following action
\bea
S = \int 
\rd^{3|4}z
\,E\, \bar \F \F
+\Big\{ \hf \int 
\rd^{3|4}z
\, E \, \frac{m}{\m}\F^2 +{\rm c.c.} \Big\}~,
\label{chiral-massive}
\eea
with $m$ a complex mass parameter. 
In the massive case $J_{(2s)}$ satisfies a more general conservation equation~\eqref{7.9b}
for some superfield $\bar T_{(2s-3)}$.
Making use of the equations of motion
\bea
-\frac{1}{4} (\cD^2-4\bar\m) \F  +\bar m \bar \F =0, \qquad
-\frac{1}{4} (\bar \cD^2-4\m) \bar \F +m \F =0,
\eea
we obtain 
\begin{subequations}
\bea
\cD_{(-1)} J_{(2s)} &=& \bar F_{(2s-1)}~, \label{7.12a}
\eea
where we have denoted 
\bea
\bar F_{(2s-1)} &=& \bar m(2s+1) \sum_{k=0}^{s-1} (-1)^{k} \binom{2s}{2k+1}
\non \\ 
&& 
\times \left\{(-1)^s +\frac{2k+1}{2s-2k+1}\right\} 
 {\cD}^k_{(2)} \bar \F \,\,{\cD}^{s-k-1}_{(2)}
 \bar \cD_{(1)} \bar \F ~.
\eea
\end{subequations}

We now  look for a superfield $\bar T_{(2s-3)}$
such that (i) it obeys the transverse antilinear constraint \eqref{7.8b}; and 
(ii) it satisfies the equation
\bea
\bar F_{(2s-1)} = s \bar A_{(2)} \bar T_{(2s-3)}~. 
\eea
Our analysis will be similar to the one performed in~\cite{BHK} in the case of four-dimensional AdS. 
We consider a general ansatz
\bea
\bar T_{(2s-3)} = \bar m \sum_{k=0}^{s-2} c_k 
{\cD}^k_{(2)} \bar \F\,
{\cD}^{s-k-2}_{(2)}
 \bar \cD_{(1)} \bar \F 
 \label{T7.15}
\eea
with some coefficients $c_k$ which have to be determined. 
For $k = 1,2,...s-2$, condition (i) implies that 
the coefficients $c_k$ must satisfy
\begin{subequations}\label{7.16}
\begin{align}
kc_k = (s-k-1) c_{s-k-1}~,\label{7.16a}
\end{align}
while (ii) gives the following equation
\begin{align}
c_{s-k-1} + s c_k + (s-1) c_{k-1} &= -\frac{2s+1}{2s} (-1)^k \binom{2s}{2k+1}
\non \\
& \qquad \qquad \times \left\{(-1)^s +\frac{2k+1}{2s-2k+1} \right\} ~.\label{7.16b}
\end{align}
Condition (ii) also implies that 
\begin{align}
(s-1) c_{s-2} +c_0 &= (2s+1)\left\{1+(-1)^s \frac{2s-1}{3}\right\}~, \label{7.16c}\\
c_0 &= -\frac{1}{s}(1+(-1)^s(2s+1))~. \label{7.16d}
\end{align}
\end{subequations}
It turns out that the equations \eqref{7.16} 
lead to a unique expression for $c_k$ given by 
\bea\label{7.17}
c_k &=& (-1)^{s+k-1} \frac{(2s+1)(s-k-1)}{2s(s-1)}
\sum_{l=0}^k \frac{1}{s-l} \binom{2s}{2l+1} \left\{ 1+(-1)^s \frac{2l+1}{2s-2l+1} \right\}  ~,~~~~  \\
&& \qquad \qquad \qquad  k=0,1,\dots s-2~. \non 
\eea

If the parameter $s$ is odd, $s=2n+1$, with  $n=1,2,\dots$, 
one can check that the equations \eqref{7.16a}--\eqref{7.16c} are identically 
satisfied. 
However, if the parameter $s$ is even, $s=2n$, with $n=1,2,\dots$, 
there appears an inconsistency: 
 the right-hand side of \eqref{7.16c} is positive, while the left-hand side 
is negative, $(s-1) c_{s-2} + c_0 < 0$. Therefore, our solution \eqref{7.17} is only consistent for $s=2n+1, n=1,2,\dots$. 

Relations \eqref{7.15}, \eqref{T7.15}, \eqref{7.16d} and \eqref{7.17} determine the non-conformal higher-spin supercurrents 
in the massive chiral model \eqref{chiral-massive}.
Unlike the conformal higher-spin supercurrents \eqref{7.15},
the non-conformal ones exist only for the odd values of $s$,
$s=2n+1$, with  $n=1,2,\dots$.


\subsection{Superconformal model with $N$ chiral superfields}

In this subsection we will generalise the superconformal model \eqref{chiral} to the case of $N$ covariantly chiral scalar superfields $\F^i$, $i=1,\dots N$,
\bea
S = \int 
\rd^{3|4}z
\,E \,{\bar \F}^i \F^i ~,
\qquad {\bar \cD}_\a \F^i = 0 ~.~
\label{Nchiral}
\eea
There exist two different types of conformal supercurrents, which are:
\bea
J^+_{(2s)} &=& S^{ij} \sum_{k=0}^s (-1)^k
\left\{ \hf \binom{2s}{2k+1} 
{\cD}^k_{(2)} \bar \cD_{(1)} \bar \F^i \,\,
{\cD}^{s-k-1}_{(2)} \cD_{(1)} \F^j  
\right. \non \\ 
&& \left.
 \qquad \qquad
+ \binom{2s}{2k} 
{\cD}^k_{(2)} \bar \F^i \,\, {\cD}^{s-k}_{(2)} \F^j \right\}~, \qquad S^{ij}= S^{ji} 
\label{J-sym}
\eea
and
\bea
J^-_{(2s)} &=& \ri \, A^{ij} \sum_{k=0}^s (-1)^k
\left\{ \hf \binom{2s}{2k+1} 
{\cD}^k_{(2)} \bar \cD_{(1)} \bar \F^i \,\,
{\cD}^{s-k-1}_{(2)} \cD_{(1)} \F^j  
\right. \non \\ 
&& \left.
 \qquad \qquad
+ \binom{2s}{2k} 
{\cD}^k_{(2)} \bar \F^i \,\, {\cD}^{s-k}_{(2)} \F^j \right\}~, \qquad A^{ij}= -A^{ji} 
\label{J-anti}
\eea
Here $S$ and $A$ are arbitrary real symmetric and antisymmetric 
 constant matrices, respectively. We have put an overall factor  $\sqrt{-1} $ in eq.~\eqref{J-anti} in order to make $J^-_{(2s)}$ real. 
One can show that the currents \eqref{J-sym} and \eqref{J-anti} are conserved on-shell:
\bea
\cD_{(-1)} J^\pm_{(2s)} = 0 \quad \Longleftrightarrow \quad 
\bar \cD_{(-1)} J^\pm_{(2s)} = 0 ~.~
\eea

The above results can be recast in terms of the matrix conformal supercurrent
$J_{(2s)} =\big(J^{ij}_{(2s)} \big)$ with components
\bea
J^{ij}_{(2s)} &:=& \sum_{k=0}^s (-1)^k
\left\{ \hf \binom{2s}{2k+1} 
{\cD}^k_{(2)} \bar \cD_{(1)} \bar \F^i \,\,
{\cD}^{s-k-1}_{(2)} \cD_{(1)} \F^j 
\right. \non \\ 
&& \left.
 \qquad \qquad
+ \binom{2s}{2k} 
{\cD}^k_{(2)} \bar \F^i \,\, {\cD}^{s-k}_{(2)} \F^j \right\}~, 
\label{520}
\eea
which is  Hermitian, $J_{(2s)}{}^\dagger = J_{(2s)}$. 
The chiral action  \eqref{Nchiral}
possesses rigid ${\rm U}(N)$ symmetry acting on the chiral column-vector $\F = (\F^i$) 
by $\F \to g \F$, with $g \in {\rm U}(N)$, which implies that 
the supercurrent \eqref{520} transforms
as $J_{(2s)} \to gJ_{(2s)} g^{-1}$.


\section{Higher-spin supercurrents for chiral matter: Integer superspin}

In this section we provide explicit realisations for the fermionic higher-spin supercurrents 
(integer superspin) in a model of a single massive chiral scalar superfield. 



We start by considering the massive action 
\bea
S = \int 
\rd^{3|4}z
 \,E\, \bar \J \J
+\Big\{\hf \int 
\rd^{3|4}z
\,E\, \frac{m}{\m} \J^2 +{\rm c.c.} \Big\}~,
\label{cmass}
\eea
where the superfield $\J$ is covariantly chiral, $\bar \cD_\a \J =0$ and $m$ is a complex mass parameter.
By a change of variables it is possible to make $m$ real. 
Let us introduce a new chiral superfield $\F$, $\bar \cD_\a \F =0$,  
related to $\J$ 
by a phase transformations,
\bea \label{phase}
\F = \re^{\ri \a/2} \J~, \qquad
m = M \re^{\ri \a}~, \quad \bar M =M~.
\eea
Then the action \eqref{cmass} turns into
\bea
S = \int 
\rd^{3|4}z
\,E \, \bar \F \F
+\Big\{ \hf\int 
\rd^{3|4}z
\,E \,  \frac{M}{\m} \F^2  +{\rm c.c.} \Big\}~.
\label{hyper2}
\eea
We emphasise that the mass parameter $M$ is now real.

In the massless case, $M=0$, 
the conserved fermionic supercurrent $J_{\a(2s-1)}$ is given by
\bea
J_{(2s-1)} &=& \sum_{k=0}^{s-1} (-1)^k 
\left\{ \binom{2s-1}{2k+1} 
{\cD}^k_{(2)}
 \cD_{(1)} \F \,\,
{\cD}^{s-k-1}_{(2)}
 \F  
\right. \non \\ 
&& \left.
 \qquad \qquad
- \binom{2s-1}{2k} 
{\cD}^k_{(2)}
  \F \,\,
{\cD}^{s-k-1}_{(2)}
\cD_{(1)} \F\right\}~.
\label{4.8}
\eea
By changing the summation index in \eqref{4.8}, it is not hard to see that $J_{(2s-1)}$ is zero for odd values of $s$. Making use of
the massless equations of motion,  $-\frac{1}{4}(\cD^2-4\bar \m)\, \F = 0$, 
one may check that $J_{(2s-1)}$ obeys, 
for $s > 1$,  the conservation equations
\bea
\cD_{(-1)} J_{(2s-1)} = 0, \qquad
\bar \cD_{(-1)} J_{(2s-1)} = 0 ~.~
\label{4.9}
\eea

We will now construct fermionic higher-spin supercurrents 
corresponding to the massive model \eqref{hyper2}.
Making use of the massive equation of motion
\bea
-\frac{1}{4} (\cD^2-4\bar\m) \F +M \bar \F =0, 
\eea
we obtain 
\bea
\cD_{(-1)} J_{(2s-1)} &=& 8Ms \sum_{k=0}^{s-1} (-1)^{k+1} \binom{2s-1}{2k} \non \\
&\times & \Big\{ {\cD}^k_{(2)} \F \,{\cD}^{s-k-1}_{(2)} \bar \F  \,\,
 + \frac{k}{2k+1} 
{\cD}^{k-1}_{(2)} \,\bar \cD_{(1)} \bar \F \,\,{\cD}^{s-k-1}_{(2)} \cD_{(1)} \F \Big\}
~.~ \label{4.10}
\eea
It can be shown that the massive supercurrent $J_{(2s-1)}$ also obeys \eqref{4.5a}. 

We now look for a superfield $T_{\a(2s-2)}$ such that (i) it obeys the longitudinal linear constraint \eqref{4.5b}; and 
(ii) it satisfies \eqref{4.7}, which is a consequence of the conservation equation \eqref{4.6}. 
For this we consider a general ansatz 
\bea
T_{(2s-2)} &=& 
\sum_{k=0}^{s-1} c_k \,{\cD}^k_{(2)} \F \,\, {\cD}^{s-k-1}_{(2)} \bar \F  \non \\
&&+ \sum_{k=1}^{s-1} d_k \,{\cD}^{k-1}_{(2)} \cD_{(1)} \F \,\, {\cD}^{s-k-1}_{(2)} \bar \cD_{(1)} \bar \F  ~.
\label{T4.11}
\eea
Condition (i) implies that the coefficients must be related by
\begin{subequations} 
\bea
c_0 = 0~, \qquad c_k = 2d_k ~, 
\label{qqq1}
\eea
while for $k=1,2, \dots s-2$, condition  (ii) gives the following recurrence relations:
\bea 
d_k + d_{k+1} &=& -\frac{8Ms}{2s-1} (-1)^{k+1} \binom{2s-1}{2k} 
 \frac{4ks+3s-1-2s^2}{(2k+1)(2k+3)} ~.
\eea
Condition (ii) also implies that
\bea
d_1 =-\frac{8}{3} Ms(s-1)~, \qquad d_{s-1} &=& -\frac{8}{2s-1} Ms(s-1)~.
\eea
\end{subequations}
The above conditions lead to a simple expression for $d_k$:
\bea
d_k &=& \frac{8Ms}{2s-1} \frac{k}{2k+1} (-1)^{k} \binom{2s-1}{2k} ~,
\eea
\label{qqq2}
where $ k=1,2,\dots s-1$ and the parameter $s$ is even 
for $J_{(2s-1)}$ to be non-zero.


\section{Concluding comments}

The constructions presented in this paper have several interesting extensions,
some of which are briefly discussed below.  

Our results can be used to construct off-shell formulations for massive higher-spin 
supermultiplets in (1,1) AdS superspace.\footnote{Two different Lagrangian 
formulations for massive higher-spin $\cN=1$ 
supermultiplets in AdS${}_3$ were developed in  \cite{KP,BSZ4}. }
This is readily achieved in the case of a half-integer superspin
by considering two dually equivalent gauge-invariant actions 
\begin{subequations} \label{9.1}
\bea
S_{\rm massive}^{\perp}&=& 
\k {S}_{\rm SCS} [ {\mathfrak H}_{(2s)}] 
+ m^{2s-1} S^{\perp}_{(s+\hf)} [{\mathfrak H}_{\a(2s)} ,
\G_{\a(2s-2)} , \bar \G_{\a(2s-2)}]~, \label{9.1a}\\
S_{\rm massive}^{\|}&=& 
\k {S}_{\rm SCS} [ {\mathfrak H}_{(2s)}] 
+m^{2s-1}S^{\|}_{(s+\hf)} [{\mathfrak H}_{\a(2s)} ,
G_{\a(2s-2)} , \bar G_{\a(2s-2)}]~. \label{9.1b}
\eea
\end{subequations}
Here the parameter $\k$ is dimensionless, while  $m$ has dimension of mass.
The superconformal action ${S}_{\rm SCS} [ {\mathfrak H}_{(2s)}] $ is obtained from
\eqref{2.34} by setting $n=2s$.
The massless actions $S^{\perp}_{(s+\hf)} $ and $S^{\|}_{(s+\hf)} $
are given by eqs. \eqref{tr-action-half} and \eqref{long-action-half}, 
respectively. In the flat-superspace limit, the actions \eqref{9.1a} and
\eqref{9.1b} reduce to those proposed in \cite{KO}.

We expect that the equations of motion in the topologically massive 
models  \eqref{9.1a} and
\eqref{9.1b} describe a subclass of the irreducible on-shell massive supermultiplets 
in (1,1) AdS superspace proposed in \cite{KNT-M}.
This is indeed the case in Minkowski superspace, as demonstrated in \cite{KO}.
However, analysis of the equations of motion in (1,1) AdS superspace 
is more complicated since we still do not have 
a closed-form expression for the higher-spin super-Cotton tensor
${\mathfrak W}_{\a(n)}$, eq. \eqref{2.43}, in terms of the prepotential 
${\mathfrak H}_{\a(n)}$ and the covariant derivatives 
$\cD_A$ of (1,1) AdS superspace. Here we simply recall the explicit structure of 
irreducible on-shell massive higher-spin supermultiplets in (1,1) AdS superspace  
\cite{KNT-M}.
Given a positive integer $n>0$, 
such a supermultiplet is realised in terms of 
a real symmetric rank-$n$ spinor $T_{\a(n)}$
constrained by 
\begin{subequations} 
\bea
\cD^\b T_{ \a_1 \cdots \a_{n-1} \b} 
= \bar \cD^\b T_{ \a_1 \cdots \a_{n-1} \b}&=&0~, \\ 
\Big( \frac{\ri}{2} \cD^\g\bar \cD_\g + m \Big) T_{\a_1 \cdots \a_n} &=&0~.
\eea
\end{subequations}
It can be shown that 
\bea 
\Big( \frac{\ri}{2} \cD^\g\bar \cD_\g\Big)^2 T_{\a_1 \cdots \a_n}  = 
\Big( \cD^a \cD_a + 2 (n + 2) |\m |^2 \Big) T_{\a_1 \cdots \a_n} \ .
\eea

New duality transformations were introduced in \cite{K16} for theories 
formulated in terms of the linearised higher-spin super-Cotton tensor
 $W_{\a(n)}$ in Minkowski superspace, eq. \eqref{eq:HSFSUniversal}.
These duality transformations can readily be generalised to arbitrary 
conformally flat backgrounds  by replacing  $W_{\a(n)}$ 
with  ${\mathfrak W}_{\a(n)}$ given by eq. \eqref{2.43}.

It is worth studying in more detail the higher-derivative Chern-Simons 
theory \eqref{2.34} on conformally flat superspace backgrounds.
It is a reducible gauge theory (following the terminology of 
the Batalin-Vilkovisky quantisation \cite{BV})
since one and the same gauge transformation \eqref{2.28} 
is generated by two gauge parameters, ${\l}_{\a(n-1)} $
and $\tilde{\l}_{\a(n-1)} $, such that their difference is longitudinal linear, 
\bea
\d_\l {\mathfrak H}_{\a(n) } =\d_{\tilde \l} {\mathfrak H}_{\a(n) } ~, 
\qquad \tilde{\l}_{\a(n-1)} := \l_{\a(n-1)}
+\bar  \cD_{(\a_1} \r_{\a_2 \dots \a_{n-1}) }~,
\eea
for arbitrary $\r_{\a(n-2)}$. It would be interesting to quantise the topological 
theory \eqref{2.34} and compute its partition function on topologically 
non-trivial backgrounds such as $S^1 \times S^2$.

Following \cite{KLT-M12}, we can introduce a real basis for the spinor covariant derivatives
which is obtained by replacing the complex operators $\cD_\a$ and $\bar \cD_\a$ with 
$\de_\a^I$, where $I ={\bf1}, {\bf2}$,
 defined by  
\bea
 \cD_\a=\frac{\re^{\ri\vf}}{\sqrt{2}}(\de_\a^{\bf 1}-\ri\de_\a^{\bf 2})~,\qquad
  \bar \cD_\a=-\frac{\re^{-\ri\vf}}{\sqrt{2}}(\de_\a^{\bf 1}+\ri\de_\a^{\bf 2})~,
 \eea 
where we have represented  $\mu=-\,\ri\,\re^{2\ri\vf} |\m|$.
The new covariant derivatives  can be shown to obey the following algebra:
\bsubeq
\bea
\{\nabla_\a^{\bf1},\de_\b^{\bf1}\}&=&
2\ri\de_{\a\b}
-4\ri |\m| M_{\a\b}
~, \qquad
\{\de_\a^{\bf2},\de_\b^{\bf2}\}=
2\ri\de_{\a\b}
+4\ri |\m| M_{\a\b}
~, \\
\{\de_\a^{\bf1},\de_\b^{\bf2}\}&=&0
~,~~~~~~~
\label{1_1-alg-AdS-2-1b}
\\
{[}\de_{a},\de_\b^{\bf1}{]}
&=&
|\m|(\g_a)_\b{}^\g\de_{\g}^{\bf1}
~, \qquad
{[}\de_{a},\de_\b^{\bf2}{]}
=
-|\m|(\g_a)_\b{}^\g\de_{\g}^{\bf2}
~, \\
{[}\de_{a},\de_b{]} &=&-4 |\m|^2 M_{ab}
~.~~~~
\label{1_1-alg-AdS-2}
\eea
\esubeq
 The graded commutation relations for the operators $\de_a$ and $\de_\a^{\bf1}$
have the following properties: 
 (i) they do not involve $\de_\a^{\bf2}$; and (ii)  they are identical
to those defining the $\cN=1$ AdS superspace, ${\rm AdS}^{3|2}$, see \cite{KLT-M12}
for the details.
 These properties mean that ${\rm AdS}^{3|2}$ is naturally  embedded in
 (1,1) AdS  superspace  as a subspace. The Grassmann variables
  $\q^\m_I =(\q^\m_{\bf 1}, \q^\m_{\bf 2} )$ may be chosen in such a way that 
 ${\rm AdS}^{3|2}$ corresponds to the surface defined by $\q^\m_{\bf 2} =0$.
 Every supersymmetric field theory in (1,1) AdS superspace 
 may be reduced to ${\rm AdS}^{3|2}$. Such $\cN=2 \to \cN=1$ AdS superspace 
 reduction may be carried out for all the higher-spin supersymmetric theories 
 constructed in this paper. Implementation of this program will be 
 described elsewhere.
 Here we only point out that 
 reducing the longitudinal model
for the massless superspin-$s$ multiplet
(presented in subsection 
\ref{subsection5.1}) to ${\rm AdS}^{3|2}$ leads to 
a new massless higher-spin gauge theory   that was not described in \cite{KP}.
Appendix \ref{AppendixB} provides the technical details of such a
reduction in the flat-superspace case.
 \\
 

\noindent
{\bf Acknowledgements:}\\
SMK is grateful to the Max Planck Institute for Gravitational Physics 
(Albert Einstein Institute) for hospitality at the final stage of this project.
The work of JH is supported by an Australian Government Research 
Training Program (RTP) Scholarship.
The work of SMK is supported in part by the Australian 
Research Council, project No. DP160103633.


\appendix 

\section{Notation, conventions and AdS identities} 
\label{AppendixA}

We follow the notation and conventions adopted in
\cite{KLT-M11}. In particular, the Minkowski metric is
$\eta_{ab}=\mbox{diag}(-1,1,1)$.
The spinor indices are  raised and lowered using
the $\rm SL(2,{\mathbb R})$ invariant tensors
\bea
\ve_{\a\b}=\left(\begin{array}{cc}0~&-1\\1~&0\end{array}\right)~,\qquad
\ve^{\a\b}=\left(\begin{array}{cc}0~&1\\-1~&0\end{array}\right)~,\qquad
\ve^{\a\g}\ve_{\g\b}=\d^\a_\b
\eea
by the standard rule:
\bea
\psi^{\a}=\ve^{\a\b}\psi_\b~, \qquad \psi_{\a}=\ve_{\a\b}\psi^\b~.
\label{A2}
\eea

We make use of real gamma-matrices,  $\g_a := \big( (\g_a)_\a{}^\b \big)$, 
which obey the algebra
\be
\gamma_a \gamma_b=\eta_{ab}{\mathbbm 1} + \varepsilon_{abc}
\gamma^c~,
\label{A3}
\ee
where the Levi-Civita tensor is normalised as
$\varepsilon^{012}=-\varepsilon_{012}=1$. The completeness
relation for the gamma-matrices reads
\be
(\gamma^a)_{\alpha\beta}(\gamma_a)^{\rho\sigma}
=-(\delta_\alpha^\rho\delta_\beta^\sigma
+\delta_\alpha^\sigma\delta_\beta^\rho)~.
\label{A4}
\ee
Here the symmetric matrices 
$(\gamma_a)^{\alpha\beta}$ and $(\gamma_a)_{\alpha\beta}$
are obtained from $\g_a=(\g_a)_\a{}^{\b}$ by the rules (\ref{A2}).
Some useful relations involving $\g$-matrices are 
\bsubeq
\bea
\ve_{abc}(\g^b)_{\a\b}(\g^c)_{\g\d}&=&
\ve_{\g(\a}(\g_a)_{\b)\d}
+\ve_{\d(\a}(\g_a)_{\b)\g}
~,
\\
\tr[\g_a\g_b\g_{c}\g_d]&=&
2\eta_{ab}\eta_{cd}
-2\eta_{ac}\eta_{db}
+2\eta_{ad}\eta_{bc}
~.
\eea
\esubeq

Given a three-vector $x_a$,
it  can be equivalently described by a symmetric second-rank spinor $x_{\a\b}$
defined as
\bea
x_{\a\b}:=(\g^a)_{\a\b}x_a=x_{\b\a}~,\qquad
x_a=-\hf(\g_a)^{\a\b}x_{\a\b}~.
\eea
In the 3D case,  an
antisymmetric tensor $F_{ab}=-F_{ba}$ is Hodge-dual to a three-vector $F_a$, 
specifically
\bea
F_a=\hf\ve_{abc}F^{bc}~,\qquad
F_{ab}=-\ve_{abc}F^c~.
\label{hodge-1}
\eea
Then, the symmetric spinor $F_{\a\b} =F_{\b\a}$, which is associated with $F_a$, can 
equivalently be defined in terms of  $F_{ab}$: 
\bea
F_{\a\b}:=(\g^a)_{\a\b}F_a=\hf(\g^a)_{\a\b}\ve_{abc}F^{bc}
~.
\label{hodge-2}
\eea
These three algebraic objects, $F_a$, $F_{ab}$ and $F_{\a \b}$, 
are in one-to-one correspondence to each other, 
$F_a \leftrightarrow F_{ab} \leftrightarrow F_{\a\b}$.
The corresponding inner products are related to each other as follows:
\bea
-F^aG_a=
\hf F^{ab}G_{ab}=\hf F^{\a\b}G_{\a\b}
~.
\eea

The Lorentz generators with two vector indices ($M_{ab} =-M_{ba}$),  one vector index ($M_a$)
and two spinor indices ($M_{\a\b} =M_{\b\a}$) are related to each other by the rules:
$M_a=\hf \ve_{abc}M^{bc}$ and $M_{\a\b}=(\g^a)_{\a\b}M_a$.
These generators 
act on a vector $V_c$ 
and a spinor $\J_\g$ 
as follows:
\bea
M_{ab}V_c=2\eta_{c[a}V_{b]}~, ~~~~~~
M_{\a\b}\J_{\g}
=\ve_{\g(\a}\J_{\b)}~.
\label{generators}
\eea

The covariant derivatives of (1,1) AdS superspace obey various identities,
which can be readily derived from 
the covariant derivatives algebra \eqref{1.2}.
We have made use of the following identities:
\begin{subequations} 
\label{1.4}
\bea 
\cD_\a\cD_\b
\!&=&\!\frac{1}{2}\ve_{\a\b}\cD^2-2{\bar \m}\,M_{\a\b}~,
\quad\qquad \,\,\,
{\bar \cD}_\a{\bar \cD}_\b
=-\frac{1}{2}\ve_{\a \b}{\bar \cD}^2+2\m\,{ M}_{\a \b}~,  \label{1.4a}\\
\cD_\a\cD^2
\!&=&\!4 \bar \m \,\cD^\b M_{\a\b} + 4{\bar \m}\,\cD_\a~,
\quad\qquad
\cD^2\cD_\a
=-4\bar \m \,\cD^\b M_{\a\b} - 2\bar \m \, \cD_\a~, \label{1.4b} \\
{\bar \cD}_\a{\bar \cD}^2
\!&=&\!4 \m \,{\bar \cD}^\b { M}_{\a \b}+ 4\m\, \bar \cD_\a~,
\quad\qquad
{\bar \cD}^2{\bar \cD}_\a
=-4 \m \,{\bar \cD}^\b {M}_{\a \b}-2\m\, \bar \cD_\a~,  \label{1.4c}\\
\left[\bar \cD^2, \cD_\a \right]
\!&=&\!4\ri \cD_{\a\b} \bar \cD^\b +6 \m\,\cD_\a = 
4\rm i \bar \cD^\b \cD_{\a\b} -6 \m\,\cD_\a~,
 \label{1.4d} \\
\left[\cD^2,{\bar \cD}_\a \right]
\!&=&\!-4\ri \cD_{\b\a}\cD^\b +6\bar \m\,{\bar \cD}_\a = 
-4\rm i \cD^\b \cD_{\b\a} +6 \bar \m\,{\bar \cD}_\a~,
 \label{1.4e}
\eea
\end{subequations} 
where $\cD^2=\cD^\a\cD_\a$, and ${\bar \cD}^2={\bar \cD}_\a {\bar \cD}^\a$. 
These relations imply the identity 
\bea
\cD^\a (\bar \cD^2- 6 \mu) \cD_\a = \cDB_\a (\cD^2 - 6 \mub) \cDB^\a ~,
\label{1.44}
\eea
which guarantees the reality of the actions considered 
in the main body of the paper.


\section{$\bm{\cN=2} \rightarrow \bm{\cN=1}$ superspace reduction}
\label{AppendixB}

In this appendix we carry out the ${\cN=2} \rightarrow {\cN=1}$ superspace 
reduction \cite{KT}  of the massless 
integer-superspin model \eqref{long-action-int}.
For simplicity our analysis is restricted to flat superspace. 
 An extension to the AdS case will be discussed elsewhere. 

In order to be consistent with the previous work \cite{KT}, 
in which the ${\cN=2} \rightarrow {\cN=1}$ superspace reduction 
of the massless half-integer-superspin models of \cite{KO} 
was studied,
we denote by 
${\mathbb D}_\a$ and $\bar {\mathbb D}_\a$ the spinor covariant 
derivatives\footnote{The operators 
${\mathbb D}_\a$ and $\bar {\mathbb D}_\a$ coincide with $D_\a$ and $\bar D_\a$
given in eq. \eqref{2.38}. However,  it is advantageous here to use
the different notation for these covariant derivatives.}
of $\cN=2$ Minkowski superspace ${\mathbb M}^{3|4}$.
They obey the 
anti-commutation relations
\bea
\{{\mathbb D}_\a, \bar {\mathbb D}_\b\}=-2\ri\, \pa_{\a\b}~,\qquad
\{{\mathbb D}_\a,{\mathbb D}_\b\}=\{ \bar {\mathbb D}_\a, \bar {\mathbb D}_\b\}=0~.
\label{N=2acd}
\eea
In order to carry out the $\cN=2 \to \cN=1$ superspace reduction, 
it is useful to introduce 
real Grassmann coordinates  $\q^\a_I$ for ${\mathbb M}^{3|4}$, 
where $I =\1, \2$. We define these coordinates 
by choosing the corresponding spinor covariant derivatives
$D^I_\a$ as in \cite{KPT-MvU}:
\bea
&&
{\mathbb D}_\a=\frac{1}{ \sqrt{2}}(D_\a^{\1}-\ri D_\a^{\2})~,\qquad
\bar {\mathbb D}_\a=-\frac{1}{ \sqrt{2}}(D_\a^{\1}+\ri D_\a^{\2})~.~~~
\label{N1-deriv}
\eea
From \eqref{N=2acd} we deduce
\bea
\big\{ D^I_\a , D^J_\b \big\} = 2{\rm i}\, \d^{IJ}  (\g^m)_{\a\b}\,\pa_m~, 
\qquad I ,J=\1, \2~.
\eea

Given an $\cN=2$ superfield $U(x, \q_I)$, we define its $\cN=1$ bar-projection
\bea
U|:= U(x, \q_I)|_{\q_{\2} =0}~,
\eea
which
is a superfield on $\cN=1$ Minkowski superspace  
${\mathbb M}^{3|2}$ parametrised
by real Cartesian coordinates $z^A= (x^a, \q^\a)$, where $\q^\a:=\q^\a_{\1}$.
The spinor covariant derivative of $\cN=1$ Minkowski superspace
$D_\a := D_\a^{\1}$
obeys the anti-commutation relation
\bea
\big\{ D_\a , D_\b \big\} = 2{\rm i}\,   (\g^m)_{\a\b}\,\pa_m~.
\eea
Finally, the ${\cN=2} \rightarrow {\cN=1}$ superspace reduction 
of the $\cN=2$ supersymmetric action is carried out using the rule
 \cite{KT}
\bea
S =\int \rd^{3|4}z \, L_{(\cN=2)}
= \int \rd^{3|2} z \, L_{(\cN=1)}~, \qquad 
L_{(\cN=1)} := -\frac{\ri}{4} (D^{\2})^2 L_{(\cN=2)}\Big|~.
\label{B.6}
\eea

Given an integer $s \geq 1$, the longitudinal formulation 
for the massless superspin-$s$ multiplet 
is realised in terms of the following dynamical variables:
\bea
\cV^{\|}_{(s)} = \Big\{\mathbb{U}_{\a(2s-2)}, \mathbb{G}_{\a(2s)}, \mathbb{\bar{G}}_{\a(2s)} \Big\} ~.
\eea
Here $\mathbb{U}_{\a(2s-2)}$ is an unconstrained real superfield, and the complex superfield $\mathbb{G}_{\a(2s)}$ is longitudinal linear,
\be
\mathbb{\bar{D}}_{(\a_1} \mathbb{G}_{\a_2 \dots \a_{2s+1})} =0~.
\ee
The dynamical superfields are defined modulo gauge transformations of the form 
\begin{subequations} \label{int-gauge-flat}
\bea
\d \mathbb{U}_{\a(2s-2)}
&=& \bar \g_{\a(2s-2)}+{\g}_{\a(2s-2)}
~, \label{U-gauge-flat} \\
\d \mathbb{G}_{\a(2s)} 
&=& \mathbb{\bar{D}}_{(\a_{1}} \mathbb{D}_{\a_{2}}\bar \g_{\a_3 \dots \a_{2s})}
~,
\eea
\end{subequations} 
where the  gauge parameter $\g_{\a(2s-2)}$ is an arbitrary transverse linear superfield,
\be
\mathbb{{\bar D}}^{\b} \g_{\b \a_1 \dots \a_{2s-3}} =0~. \label{ee1}
\ee
The gauge-invariant action is 
\bea
S_{(s)}^{\|}
&=& \Big(-\hf\Big)^{s}\int 
\rd^{3|4}z \,
\bigg\{\frac{1}{8}\mathbb{U}^{\a(2s-2)}\mathbb{D}^{\g}{\mathbb{\bar D}}^{2}\mathbb{D}_{\g} \mathbb{U}_{\a(2s-2)}
\non \\
&&+\frac{s}{2s+1} \mathbb{U}^{\a(2s-2)}\Big(\mathbb{D}^{\b} \mathbb{{\bar D}}^{\g}
\mathbb{G}_{\b \g \a(2s-2) }
-\mathbb{\bar D}^{\b}\mathbb{D}^{\g} \mathbb{\bar{G}}_{\b \g \a(2s-2)} \Big) \non \\
&&+\frac{s}{2s-1}  \mathbb{\bar{G}}^{\a(2s)} \mathbb{G}_{\a(2s)}
+ \frac{s}{2(2s+1)}\Big(\mathbb{G}^{\a(2s)} \mathbb{G}_{\a(2s)}+ \mathbb{\bar{G}}^{\a(2s)} \mathbb{\bar G}_{\a(2s)}\Big) 
\bigg\}~.
\label{long-action-flat}
\eea

Making use of the representation \eqref{N1-deriv}, the transverse linear constraint \eqref{ee1} takes the form
\be 
D^{\underline{2} \, \b} \g_{\b \a_1 \dots \a_{2s-3}} 
= \ri D^{\underline{1} \b} \g_{\b \a_1 \dots \a_{2s-3}}~.
\ee
It follows that $\g_{\a(2s-2)}$ has 
two independent $\theta_{\underline{2}}$-components, which are:
\bea
\g_{\a(2s-2)} |, \qquad D^{\underline{2}}_{(\a_1} \g_{\a_2 \dots \a_{2s-1})}|~.
\eea
The gauge transformation of $\mathbb{U}_{\a(2s-2)}$, eq. \eqref{int-gauge-flat}, 
allows us to impose two conditions
\bea
\mathbb{U}_{\a(2s-2)} | =0~, \qquad D^{\underline{2}}_{(\a_1} \mathbb{U}_{\a_2 \dots \a_{2s-1})} | =0~.
\label{q1}
\eea
In this gauge we define the following unconstrained real ${\cN=1}$ superfields:
\begin{subequations} \label{q2}
\bea
U_{\a(2s-3)} &:=& \frac{\ri}{s} D^{\underline{2} \b} \mathbb{U}_{\b \a(2s-3)} | ~, \label{U1} \\
U_{\a(2s-2)} &:=& -\frac{\ri}{4s}(D^{\underline{2}})^2 \mathbb{U}_{\a(2s-2)}|~. \label{U2}
\eea
\end{subequations}
The residual gauge freedom, which preserves the gauge conditions \eqref{q1}, 
is described by unconstrained real ${\cN=1}$ superfield 
parameters $\z_{\a(2s-2)}$ and $\l_{\a(2s-1)}$ defined by 
\begin{subequations}
\bea
\g_{\a(2s-2)} | &=& \frac{\ri}{2} \z_{\a(2s-2)}~, \qquad {\bar \z}_{\a(2s-2)} = \z_{\a(2s-2)}~, \label{q3.a} \\
D^{\underline{2}}_{(\a_1} \g_{\a_2 \dots \a_{2s-1})} | &=& \hf \l_{\a(2s-1)} ~, \qquad {\bar \l}_{\a(2s-1)} = \l_{\a(2s-1)}~, \label{q3.b}
\eea
\end{subequations}
The gauge transformation laws of the superfields \eqref{q2} are
\begin{subequations}
\bea
\d U_{\a(2s-3)} &=& -\frac{\ri}{s} D^{\b} \z_{\b \a(2s-3)}~, \label{q4.a}\\
\d U_{\a(2s-2)} &=&  \frac{1}{2s} D^{\b} \l_{\b \a(2s-2)} \label{q4.b}~,
\eea
\end{subequations}

We now turn to reducing $\mathbb{G}_{\a(2s)}$ to ${\cN=1}$ superspace. 
From the point of view of ${\cN=1}$ supersymmetry, 
$\mathbb{G}_{\a(2s)}$ is equivalent to two unconstrained complex  superfields, 
which we define as follows: 
\begin{subequations} \label{G-comp}
\bea
\mathbb{G}_{\a(2s)}| &=& -\hf(G_{\a(2s)} + \ri H_{\a(2s)})~, \\
\ri D^{\underline{2} \,\b} \mathbb{G}_{\b \a(2s-1)} | &=& \Phi_{\a(2s-1)} + \ri \Psi_{\a(2s-1)}~.
\eea
\end{subequations}
Making use of the gauge transformation \eqref{int-gauge-flat} gives
\begin{subequations} \label{G-gauge2}
\bea
\d \mathbb{G}_{\a(2s)}
&=& -\ri \pa_{(\a_1 \a_2} {\bar \g}_{\a_3 \dots \a_{2s})} + \ri D^{\underline{1}}_{(\a_1} D^{\underline{2}}_{\a_2} {\bar \g}_{\a_3 \dots \a_{2s})}
~, \label{U-gauge-flat2} \\
\ri D^{\underline{2} \,\b} \d \mathbb{G}_{\b \a(2s-1)}
&=& \ri \Big\{-\ri \frac{2s-1}{2s} \pa^{\b}\,_{\a_1} D^{\underline{2}}_{(\b} {\bar \g}_{\a_2 \dots \a_{2s-1})} \non \\
\qquad &+& \frac{s-1}{s} \pa_{(\a_1 \a_2} D^{\b} {\bar \g}_{\a_3 \dots \a_{2s-1}) \b} -2 D^{\b} \pa_{\b(\a_1} {\bar \g}_{\a_2 \dots \a_{2s-1})}\non \\
\qquad &+& \frac{2s+1}{4s} D^2 D^{\underline{2}}_{(\a_1} {\bar \g}_{\a_2 \dots \a_{2s-1})} \Big\}
~,
\eea
\end{subequations} 
At this stage one should recall that upon imposing the ${\cN=1}$ supersymmetric gauge conditions \eqref{q1} the residual gauge freedom is described by the gauge parameters \eqref{q3.a} and \eqref{q3.b}. From \eqref{G-gauge2} we read off the gauge transformations of the ${\cN=1}$ complex superfields \eqref{G-comp}
\begin{subequations} \label{N1-gauge}
\bea
\d \mathbb{G}_{\a(2s)} |
&=& -\hf \Big\{\pa_{(\a_1 \a_2} \z_{\a_3 \dots \a_{2s})} +  \ri D_{(\a_1} \l_{\a_2 \dots \a_{2s})}\Big\}
~, \\
\ri D^{\underline{2} \,\b} \d \mathbb{G}_{\b \a(2s-1)} |
&=& - \frac{2s-1}{4s} \pa^{\b}\,_{(\a_1} {\l}_{\a_2 \dots \a_{2s-1})\b} 
- \ri \frac{2s+1}{8s} D^2 \l_{\a(2s-1)} \non \\
\qquad &+& \frac{s-1}{2s} \pa_{(\a_1 \a_2} D^{\b} {\z}_{\a_3 \dots \a_{2s-1}) \b} - D^{\b} \pa_{\b(\a_1} {\z}_{\a_2 \dots \a_{2s-1})}
~.
\eea
\end{subequations} 

In the ${\cN=1}$ supersymmetric gauge \eqref{q1}, 
$\mathbb{U}_{\a(2s-2)}$ is described by two unconstrained real superfields $U_{\a(2s-3)}$ and $U_{\a(2s-2)}$ defined according to \eqref{q2}, and their gauge transformation laws are given by eqs. \eqref{q4.a} and \eqref{q4.b}, respectively. It follows from the gauge transformations \eqref{q4.a}, \eqref{q4.b} and \eqref{N1-gauge} that in fact we are dealing with two different gauge theories. One of them is formulated in terms of the  unconstrained real gauge superfields
\bea
\{G_{\a(2s)}, U_{\a(2s-3)}, \J_{\a(2s-1)}\}~,
\label{B.21}
\eea
which are defined modulo gauge transformations of the form
\begin{subequations} \label{type1-gauge}
\bea
\d G_{\a(2s)} 
&=& \pa_{(\a_1 \a_2} \z_{\a_3 \dots \a_{2s})} ~, \label{B.22a}\\
\d U_{\a(2s-3)} &=& -\frac{\ri}{s} D^{\b} \z_{\b \a(2s-3)}~, \label{B.22b}\\
\d \J_{\a(2s-1)} 
&=& -\ri \frac{s-1}{2s} \pa_{(\a_1 \a_2} D^{\b} {\z}_{\a_3 \dots \a_{2s-1}) \b} +\ri D^{\b} \pa_{\b(\a_1} {\z}_{\a_2 \dots \a_{2s-1})}
~,
\eea
\end{subequations} 
where the gauge parameter $\z_{\a(2s-2)}$ is unconstrained real. 
The other theory is described by the gauge superfields 
\bea
\{H_{\a(2s)}, U_{\a(2s-2)}, \Phi_{\a(2s-1)}\}
\label{B.23}
\eea
with the following gauge freedom 
\begin{subequations} \label{type2-gauge}
\bea
\d H_{\a(2s)}
&=&  D_{(\a_1} \l_{\a_2 \dots \a_{2s})}~, \label{B.24a} \\
\d U_{\a(2s-2)} &=&  \frac{1}{2s}D^\b \l_{\b \a(2s-2)} ~, \label{B.24b} \\
\d \Phi_{\a(2s-1)}
&=& - \frac{1}{8s} \Big\{(4s-2) \pa^{\b}\,_{(\a_1} {\l}_{\a_2 \dots \a_{2s-1})\b} 
+ \ri (2s+1) D^2 \l_{\a(2s-1)}\Big\}
~.
\eea
\end{subequations}

Applying the reduction rule \eqref{B.6}
to the action \eqref{long-action-flat} gives two decoupled $\cN=1$ supersymmetric
actions, which are described  in terms of the dynamical variables
\eqref{B.21} and \eqref{B.23}, respectively. 
In the former case, the superfield $\J_{\a(2s-1)}$ is  auxiliary.
Integrating it out, we arrive at the following action:
\bea
S
&=& -\Big(-\hf\Big)^{s} \,\frac{s^2(s-1)}{2s-1} \frac{\ri}{2}\int 
\rd^{3|2}z \,
\bigg\{ 
 \frac{1}{2s} G^{\a(2s)} D^2 G_{\a(2s)}
 \non \\
&& 
-
\frac{\ri}{s-1} G^{\a(2s-1) \b} \pa_{\b}\, ^{\g} G_{\a(2s-1) \g}
-2\ri U^{\a(2s-3)} \pa^{\b \g} D^{\d} G_{\b \g \d \a(2s-3)}
 \non \\
&&+2 \,U^{\a(2s-3)} \Box U_{\a(2s-3)} 
+  \frac{(2s-3)(s-2)}{2s-1} \, \pa_{\d \l} U^{\d \l \a(2s-5)} \pa^{\b \g} U_{\b \g \a(2s-5)} \non \\
&&- \hf \,\frac{2s-3}{2s-1} D_\b U^{\a(2s-4)\b} D^2 D^\g U_{\g \a(2s-4)} 
 \bigg\}~.
\label{type1-action}
\eea
This action is invariant under  the gauge transformations
\eqref{B.22a} and \eqref{B.22b}.

In the latter case, the superfield $\Phi_{\a(2s-1)}$ is  auxiliary. 
Integrating it out, we obtain the following gauge-invariant action: 
\bea
S
&=& \Big(-\hf\Big)^{s} \frac{s }{2s-1} \ri \int 
\rd^{3|2}z \,
\bigg\{ 
\frac{1}{2} H^{\a(2s)} D^2 H_{\a(2s)} 
+\ri H^{\a(2s-1) \b} \pa_{\b} \,^{\g} H_{\a(2s-1) \g} \non \\
&&+ 2 \ri (2s-1)U^{\a(2s-2)} \pa^{\b \g} H_{\b \g \a(2s-2)}
+ (2s-1 ) U^{\a(2s-2)} D^2 U_{\a(2s-2)}
\non \\
&& 
+ 2 (2s-1)(s-1) D_{\b} U^{\b \a(2s-3)} D^{\g} U_{\g \a(2s-3)} 
\bigg\}~.
\label{type2-action}
\eea
This action is invariant under  the gauge transformations
\eqref{B.24a} and \eqref{B.24b}. Modulo an overall normalisation factor, 
\eqref{type2-action} coincides with 
the off-shell $\cN=1$ supersymmetric action 
for massless superspin-$s$ multiplet \cite{KT}
in the form given in \cite{KP}.

The action \eqref{type1-action} defines a new $\cN=1$ supersymmetric 
higher-spin theory which did not appear in the analysis of \cite{KT}.
It may be shown that at the component level it reduces, upon imposing 
a Wess-Zumino gauge and eliminating the auxiliary fields, 
to a sum of two massless actions, one of which is the bosonic Fronsdal-type  
spin-$s$ model and the other is the fermionic Fang-Fronsdal-type 
spin-$(s+\hf)$ model. 


\begin{footnotesize}

\end{footnotesize}

\end{document}